\documentclass[11pt,a4paper]{article}
 
\usepackage{amsmath,amsthm,amssymb}
\usepackage{graphics,graphicx}
\usepackage{epsfig}
\usepackage{multicol}
\usepackage{color}
\usepackage{mathrsfs}
\makeatletter
\@addtoreset{equation}{section}
\makeatother

\usepackage{braket}

\begin{document}

\begin{flushright}
\end{flushright}
\begin{center}
\LARGE{\bf 
Does Compton/Schwarzschild duality in higher dimensions exclude TeV quantum gravity?}
\end{center}

\begin{center}
\large{\bf Matthew J. Lake} ${}^{a,b}$\footnote{mjlake@ntu.edu.sg}\large{\bf and Bernard Carr} ${}^{c,d}$\footnote{B.J.Carr@qmul.ac.uk} 
\end{center}
\begin{center}
\emph{ ${}^a$ School of Physics, Sun Yat-Sen University, Guangzhou 510275, China\\}
\emph{ ${}^b$ School of Physical and Mathematical Sciences, Nanyang Technological University, 637371 Singapore, Singapore \\}
\emph{ ${}^c$ School of Physics and Astronomy, Queen Mary University of London, Mile End Road, London E1 4NS, U.K. \\}
\emph{ ${}^d$Research Center for the Early Universe (RESCEU),
Graduate School of Science, \\ University of Tokyo, Tokyo 113-0033, Japan} 
\vspace{0.1cm}
\end{center}



\begin{abstract}
In three spatial dimensions, the Compton wavelength $(R_C \propto M^{-1}$) and Schwarzschild radius $(R_S \propto M$) are dual under the transformation $M \rightarrow M_{P}^2/M$, where $M_{P}$ is the Planck mass. This suggests that there could be a fundamental link -- termed the Black Hole Uncertainty Principle or Compton-Schwarzschild correspondence -- between elementary particles with $M < M_{P}$ and black holes in the $M > M_{P}$ regime. In the presence of $n$ extra dimensions, compactified on some scale $R_E$ exceeding the Planck length $R_P$, one expects $R_S \propto M^{1/(1+n)}$ for $R_P < R < R_E$, which breaks this duality. However, it may be restored in some circumstances because the {\it effective} Compton wavelength of a particle depends on the form of the $(3+n)$-dimensional wavefunction.  If this is spherically symmetric, then one still has $R_C \propto M^{-1}$, as in the $3$-dimensional case. The effective Planck length is then increased and the Planck mass reduced, allowing the possibility of TeV quantum gravity and black hole production at the LHC. However, if the wave function of a particle
is asymmetric and has a scale $R_E$ in the extra dimensions, then $R_C \propto M^{-1/(1+n)}$, so that the duality between $R_C$ and $R_S$ is preserved. In this case, the effective Planck length is increased even more but the Planck mass is unchanged, so that TeV quantum gravity is precluded and black holes cannot be generated in collider experiments. Nevertheless, the extra dimensions could still have consequences for the detectability of black hole evaporations and the enhancement of pair-production at accelerators on scales below $R_E$. Though phenomenologically general for higher-dimensional theories, our results are shown to be consistent with string theory via the minimum positional uncertainty derived from $D$-particle scattering amplitudes.
\end{abstract}



\section{Introduction} \label{Sec.1}

\begin{figure} \label{MR}
   \begin{center}
   \psfig{file=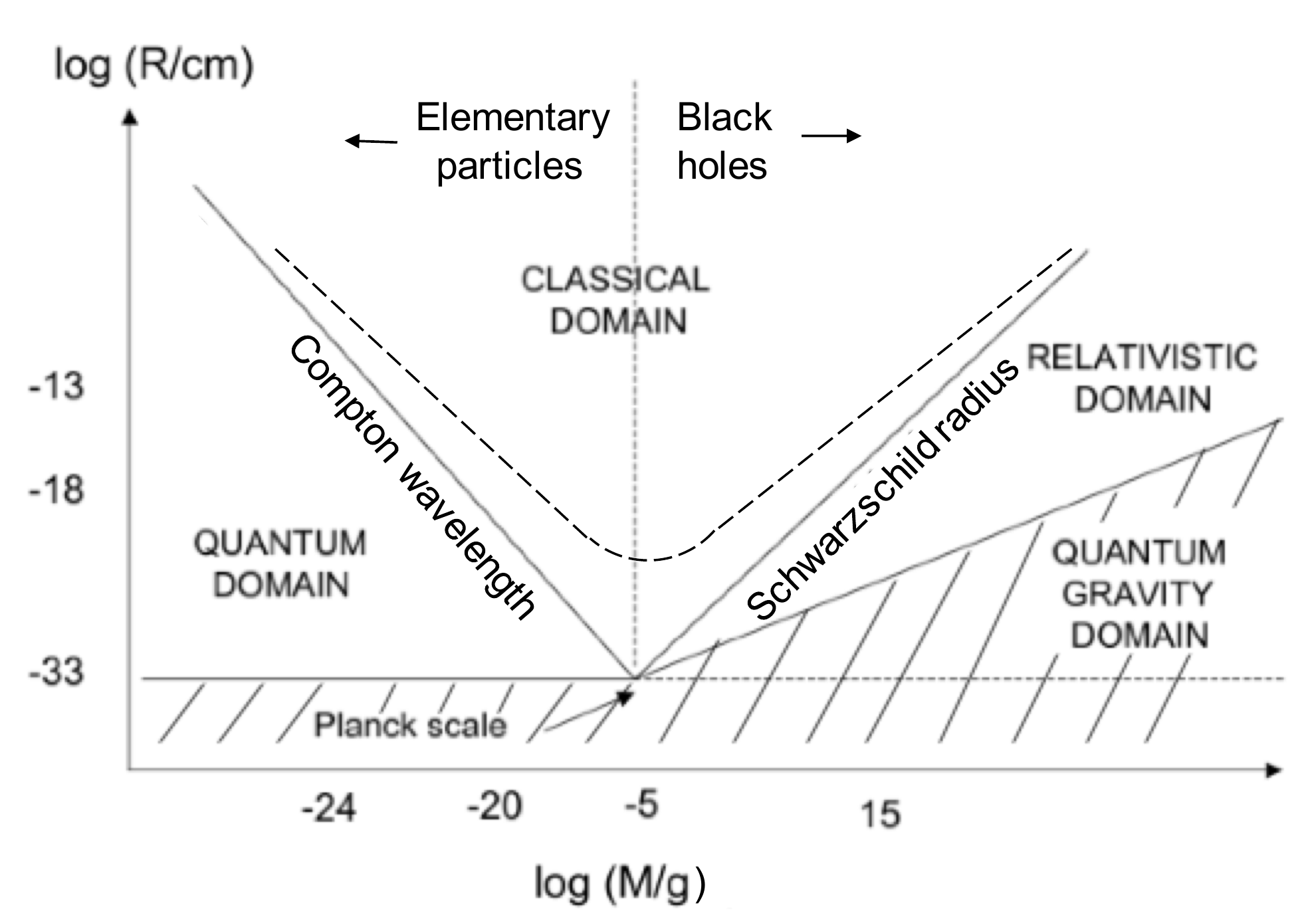,width=3.5in}
    \caption{This shows the division of the ($M,R$) diagram into different physical domains. 
Also shown are  the Compton and Schwarzschild radii and the lines corresponding to 
    the Planck mass, length and density. The broken  line gives a smooth transition between the Compton and Schwarzschild lines, as postulated by the BHUP correspondence. }     
    \end{center}
\end{figure}

A key feature of the microscopic domain is the (reduced) Compton wavelength for a particle of rest mass $M$, which is $R_C = \hbar/(Mc)$. In the $(M,R)$ diagram of Fig.~\ref{MR}, the region corresponding to $R<R_C$ might be regarded as the `quantum domain' in the sense that the classical description breaks down there. A key feature of the macroscopic domain is the Schwarzschild radius for a body of mass $M$, which corresponds to the size of a black hole of this mass and is $R_S = 2GM/c^2$. The region $R<R_S$ might be regarded as the `relativistic domain' in the sense that there is no stable classical configuration in this part of Fig.~\ref{MR}. 

Despite being essentially relativistic results, it is interesting that both these expressions can be derived from a semi-Newtonian treatment in which one invokes a maximum velocity $c$ but no other relativistic effects  \cite{Lake:2016did}. The Compton line can be derived from the Heisenberg Uncertainty Principle (HUP), which requires that the uncertainty in the position and momentum of a particle satisfy $\Delta x \gtrsim \hbar/\Delta p$, by arguing that the momentum of a particle of mass $M$ is bounded by $Mc$. This implies that  one cannot localize it on a scale less than $\hbar/(Mc)$ and is equivalent to substituting $\Delta x \rightarrow R$ and $\Delta p \rightarrow Mc$ in the uncertainty relation. Later, we discuss more rigorous ways of determining the Compton scale, in both relativisitc and non-relativistic quantum theory, though there is always some ambiguity in the precise numerical coefficient. 
The expression for the Schwarzschild radius is derived rigorously from general relativity but exactly the same expression can be obtained by equating the escape velocity in Newtonian gravity to $c$. 

The Compton and Schwarzschild lines intersect at around the Planck scales,
\begin{eqnarray} \label{planck}
R_{P} = \sqrt{ \hbar G/c^3} \simeq 10^{-33} \mathrm {cm} \, , \quad
M_{P} = \sqrt{ \hbar c/G} \simeq 10^{-5} \mathrm g \, ,
\end{eqnarray}
and naturally divide the $(M,R)$ diagram in Fig.~\ref{MR} into three domains, which for convenience we label quantum, relativistic and classical. 
There are several other interesting lines in Fig.~\ref{MR}. The vertical line $M=M_{P}$ marks the division between elementary particles ($M <M_{P}$) and black holes ($M > M_{P}$), since the event horizon of a black hole is usually required to be larger than the Compton wavelength associated with its mass. The horizontal line $R=R_{P}$ is significant because quantum fluctuations in the metric should become important below this \cite{wheeler}. Quantum gravity effects should also be important whenever the density exceeds the Planck value, $\rho_{P} = c^5/(G^2  \hbar) \simeq 10^{94} \mathrm {g \, cm^{-3}}$, corresponding to the sorts of curvature singularities associated with the big bang or the centres of black holes \cite{CaMoPr:2011}. This implies $R < R_{P}(M/M_{P})^{1/3}$, which is well above the $R = R_{P}$ line in Fig.~\ref{MR} for $M \gg M_P$, so one might regard the shaded region as specifying the `quantum gravity' domain. This point has been invoked to support the notion of Planck stars \cite{rovelli} and could have important implications for the detection of evaporating black holes \cite{barrau}.   

The Compton and Schwarzschild lines transform into one another under the substitution $M \rightarrow M_{P}^2/M$, corresponding to a reflection in the line $M = M_{P}$ in Fig.~\ref{MR}. This interchanges sub-Planckian and super-Planckian mass scales and suggests some connection between elementary particles and black holes. The lines also transform into each other under the transformation $R \rightarrow R_{P}^2/R$, corresponding to a reflection in the line $R = R_{P}$. This turns super-Planckian length scales into sub-Planckian ones, which might be regarded as unphysical. However, we note that each line maps into itself under the combined T-duality transformation 
\begin{eqnarray} \label{T-duality}
M \rightarrow M_{P}^2/M, \ \ \ R \rightarrow R_{P}^2/R \, .
\end{eqnarray} 
T-dualities arise naturally in string theory and are known to map momentum-carrying string states to winding states and vice-versa \cite{Zwiebach:2004tj}. In addition, since they map sub-Planckian length scales to super-Planckian ones, this  allows the description of physical systems in an otherwise inaccessible regime \cite{Green:1987sp,Polchinski:1998rq}. 

Although the Compton and Schwarzschild boundaries correspond to straight lines in the logarithmic plot of Fig.~\ref{MR}, this form presumably breaks down near the Planck point due to quantum gravity effects. One might envisage two possibilities: either there is a smooth minimum, as indicated by the broken line in Fig.~\ref{MR}, so the Compton and Schwarzschild lines in some sense merge, or there is some form of phase transition  or critical point at the Planck scale, so that the separation between particles and black holes is maintained. Which alternative applies has important implications for the relationship between elementary particles and black holes \cite{CaMuNic:2014}. This may link to the issue of T-duality since this could also play a fundamental role in relating point particles and black holes.
 
One way of obtaining a smooth transition between the Compton and Schwarzschild lines is to invoke some connection between the uncertainty principle on microscopic scales and black holes on macroscopic scales. This is termed the Black Hole Uncertainty Principle (BHUP) correspondence \cite{Ca:2014} and also the Compton-Schwarzschild correspondence when discussing an interpretation in terms of extended de Broglie relations \cite{Lake:2015pma}. It is manifested in a unified expression for the Compton wavelength and Schwarzschild radius. 
The simplest expression of this kind would be 
\begin{equation} \label{BHUP}
R_{CS} =  \frac{\beta \hbar}{Mc} + \frac{2GM}{c^2}    \, ,
\end{equation}
where $\beta$ is the (somewhat arbitrary) constant appearing in the Compton wavelength. In the sub-Planckian regime this can be interpreted as a modified Compton wavelength:
\begin{equation} \label{GUP2}
R_C'  = \frac{\beta \hbar}{Mc} \left[1  + \frac{2}{\beta}  \left( \frac{M}{M_P} \right)^2 \right]  \quad (M \ll M_P)  \, , 
\end{equation}
with the second term corresponding to a small correction of the kind invoked by the Generalised Uncertainty Principle  \cite{Adler1}. In the super-Planckian regime, it can be interpreted as a modified Schwarzschild radius:
\begin{equation} \label{GEH1} 
R_S' = \frac{2GM}{c^2} \left[ 1 + \frac{\beta}{2} \left(\frac{M_P}{M} \right)^2 \right] \quad (M \gg M_P) \, ,
\end{equation}
with the second term corresponding to a small correction to the Schwarzschild expression; this has been termed the Generalised Event Horizon \cite{Ca:2014}. 
More generally, the BHUP correspondence might allow any unified expression $R_{CS}(M)$ which has the asymptotic behaviour $\beta \hbar/(Mc)$ for $M \ll M_{P}$ and $2GM/c^2$ for $M \gg M_{P}$. One could envisage many such expressions but we are particularly interested in those which -- like Eq.~(\ref{BHUP}) -- are dual under under the transformation $M \rightarrow M_{P}^2/M$. The considerations of this paper are not dependent on the validity of the BHUP correspondence itself but we mention this as an example of a particular context in which the duality arises.

The black hole boundary in Fig.~\ref{MR} assumes there are three spatial dimensions but many theories suggest that dimensionality could increase on small scales. In particular, superstring theory is consistent only in $(9+1)$ spacetime dimensions, even though our observable universe is $(3+1)$-dimensional. In current models, ordinary matter is described by open strings, whose end-points are confined to a $(p+1)$-dimensional $D_p$-brane, while gravity is described by closed strings that propagate in the bulk \cite{Zwiebach:2004tj,Green:1987sp,Polchinski:1998rq}. In the Randall-Sundrum picture \cite{rs}, $p=3$ and one of the extra dimensions is large (i.e. much larger than the Planck scale), so the universe corresponds to a $D_3$ brane in a $5$-dimensional bulk. The bulk dimension is usually viewed as being warped in an anti-de Sitter space, so that the $D_3$-brane has some finite thickness, and this is equivalent to having a compactifed extra dimension. 
One could also consider models with more than one large dimension and
this might be compared to the model of Arkani-Hamed et al. \cite{arkani}, in which there are $n$ extra spatial dimensions, all compactified on the same scale. 
One could also consider models with a hierarchy of  compacitifed dimensions,
so that
the dimensionality of the universe  increases as one goes to smaller scales. 
 
This motivates us to consider the behavior of black holes and quantum mechanical particles in spacetimes with extra directions. For simplicity, we initially assume that all the  extra dimensions in which matter is free to propagate are compactified on a single length scale $R_E$, corresponding to a mass-scale $M_E \equiv \hbar/(c R_E)$. If there are $n$ extra dimensions, and black holes with $R_S < R_E$ are assumed to be approximately spherically symmetric with respect to the full $(3+n)$-dimensional space, then the Schwarzschild radius scales as $M^{1/(1+n)}$ rather than $M$ for $M <  c^2R_E/G = M_P^2/M_E$ \cite{kanti2004},
so the slope of the black hole boundary in Fig.~\ref{MR} becomes shallower in this range of $M$. The question now arises of whether the $M$ dependence of $R_C$ is also affected by the extra dimensions. The usual assumption is that it is not, so that one still has $R_C \propto M^{-1}$. In this case, the intersect of the Schwarzschild and Compton lines becomes
\begin{equation} \label{revisedplanck}
R_{P}' \simeq (R_{P}^2R_E^n)^{1/(2+n)}, \quad   
M_{P}' \simeq (M_{P}^2M_E^n)^{1/(2+n)} \,.
 \end{equation}
This gives $M_{P}' \simeq M_{P}$ and $R_{P}' \simeq R_{P}$ for $R_E \simeq R_{P}$ but $M_{P}'  \ll M_{P}$ and $R_{P}'  \gg R_{P}$ for $R_E  \gg R_{P}$. As is well known, the higher-dimensional Planck mass therefore decreases (allowing the possibility of TeV quantum gravity) and the higher-dimensional Planck length increases \cite{arkani}. 

In principle, such effects would permit the production of small black holes at the Large Hadron Collider (LHC), with their evaporation leaving a distinctive signature \cite{dimo,giddings,anchor}. However, there is still no evidence for this \cite{atlas}, which suggests that either the extra dimensions do not exist or they have a compactification scale $R_E$ which is so small that $M_P'$ exceeds the energy attainable by the LHC.  In this paper we point out another possible reason for the failure to produce black holes at accelerators. We argue that in some circumstances one expects $R_C$ to scale as $M^{-1/(1+n)}$ rather than $M^{-1}$. This has the attraction that it preserves the T-duality between $R_C$ and $R_S$; later we present arguments for why one might expect this.
In this case, Eq.~(\ref{revisedplanck}) no longer applies. Instead, the higher-dimensional Planck mass is unchanged but the Planck length is increased to
\begin{equation} \label{R_*}
R_{*} \simeq
(R_{P} R_E^n)^{1/(1+n)} \, ,
\end{equation}
which is even larger than before. While there is no TeV quantum gravity in this scenario, we will see that the preservation of duality has interesting physical implications.

The plan of the paper is as follows. Sec.~\ref{Sec.2} considers the derivation of the standard expression for the 3D Compton wavelength. Sec.~\ref{Sec.3} discusses the (well-known) expression for the Schwarzschild radius for a $(3 + n)$-dimensional black hole. Sec.~\ref{Sec.4} discusses the form of the Uncertainty Principle in higher dimensions, emphasizing that this depends crucially on the form assumed for the wave function in the higher-dimensional space. Sec.~\ref{Sec.5} then derives the associated expressions for the effective Compton wavelength in higher dimensions. Sec.~\ref{Sec.6} explores the consequences of our claim for the detectability of primordial black hole evaporations and recent $D$-particle scattering results. Sec.~\ref{Sec.7} draws some general conclusions and suggests  future work. 

\section{Derivations of the 3-dimensional Compton wavelength } \label{Sec.2}

The Compton wavelength is defined as $R_C = h /(Mc)$ and first appeared historically in the expression for the Compton cross-section in the scattering of photons off electrons \cite{Rae00}. Subsequently, it has arisen in various other contexts.  For example, it is relevant to processes which involve turning photon energy ($hc/\lambda$) into rest mass energy ($Mc^2$) and the {\it reduced} Compton wavelength $\hbar /(Mc)$ appears naturally in the Klein-Gordon and Dirac equations. 
One can also associate the Compton wavelength with the {\it localisation} of a particle and this is most relevant for the considerations of this paper. There are both non-relativistic and relativistic arguments for this notion,   
so we will consider these in turn. It is important to distinguish these diffferent contexts when discussing how the expression for the Compton wavelength is modified in higher-dimensional models but in this section we confine attention to the 3-dimensional case.

We first consider a non-relativistic argument which combines the de Broglie relations, 
\begin{equation} \label{deBroglie2}
E = \hbar\omega,  
\quad
\vec{p} = \hbar\vec{k} \,, 
\end{equation}
with the non-relativistic expression for the $3$-momentum $\vec{p} = M\vec{v}$
and a maximum speed $|\vec{v}| < c$. Then Eq.~(\ref{deBroglie2}) gives
\begin{eqnarray} \label{k^2_bound}
k = |\vec{k}| < \frac{Mc}{\hbar} \quad \Rightarrow \quad \lambda = \frac{2 \pi}{k} > \frac {2 \pi \hbar}{Mc} \quad \Rightarrow \quad R_C = \frac{h}{Mc} \, .
\end{eqnarray}
Though the numerical factors in this argument are imprecise, detailed calculations in quantum field theory and compelling observational evidence \cite{Berestetsky:1982aq} suggest this result is at least qualitatively correct.
This argument can be related to the Uncertainty Principle if Eq.~(\ref{k^2_bound}) is viewed as giving an upper bound on the wave-number of the momentum operator eigenfunctions, or equivalently a lower bound on the de Broglie wavelength, such that:
\begin{eqnarray} \label{ident}
(\Delta p_x)_{\rm max} \simeq M c \, , \ \ \ (\Delta x)_{\rm min} \simeq R_C \, .
\end{eqnarray}  
As discussed in Appendix  A, 
one must distinguish between uncertainties in $x$ and $p_x$ associated with unavoidable noise in the measurement process and the standard deviation associated with repeated measurements which do not disturb the system prior to wave function collapse. 
In the latter case, one often uses the notation $\Delta_{\psi}$ to stress the dependence on the wave vector 
$\Ket{\psi}$. Throughout this paper, we refer to uncertainties in the latter sense but drop the subscript $\psi$ for convenience. 

We now present an 
alternative non-relativistic argument for identifying the maximum possible uncertainty in the momentum $(\Delta p_x)_{\rm max}$ with the rest mass of the particle in order to obtain a minimum value of the position uncertainty $(\Delta x)_{\rm min} \simeq R_C$. 
Mathematically, this can be achieved by defining position and momentum operators, $\hat{\vec{r}}$ and $\hat{\vec{p}}$, and their eigenfunctions in the position space representation, in the usual way,
\begin{eqnarray} \label{r}
\hat{\vec{r}} = \vec{r}, \quad
\phi(\vec{r}',\vec{r}) = \delta(\vec{r}-\vec{r}'); \quad
\hat{\vec{p}} = -i\hbar \vec{\nabla}, \quad
\phi(\vec{k},\vec{r}) = e^{i\vec{k}.\vec{r}},
\end{eqnarray} 
and then introducing an infrared cut-off in the expansion for $\psi(\hat{\vec{r}})$ in terms of $\phi^{*}(\vec{k},\vec{r})$ or for $\psi(\vec{k})$ in terms of $\phi(\vec{k},\vec{r})$:
\begin{eqnarray} \label{r-space_exp}
\psi(\vec{r}) = \int_{0}^{Mc/\hbar} \psi(\vec{k}')e^{-i\vec{k}'.\vec{r}}d^3k', \ \ \ 
\psi(\vec{k}) = \int_{h/(Mc)}^{\infty} \psi(\vec{r}')e^{i\vec{k}.\vec{r}'}d^3r' \, . 
\end{eqnarray} 
 While $\psi$ is normalisable but not an eigenstate of $\vec{r}$ and $\vec{p}$, $\phi$ is an eigenstate but non-normalisable. In the momentum space representation, $\hat{\vec{r}}$ and $\hat{\vec{p}}$ and their eigenfunctions take the form 
\begin{eqnarray} \label{r*}
\hat{\vec{r}} = -i\hbar \vec{\nabla}, \quad
\phi(\vec{k},\vec{r}) = e^{i\vec{k}.\vec{r}}; \quad
\hat{\vec{p}} = \vec{p},
\quad
\phi(\vec{k}',\vec{k}) = \delta(\vec{k}-\vec{k}'),
\end{eqnarray}
and consistency requires us to introduce an ultraviolet cut-off in $k$ at
$k_{\rm max} = Mc/\hbar$. Ths implies an infrared cut-off, $r_{\rm min} = h/(Mc)$, 
so that the extension of $\psi(\vec{r})$ in position space is bounded from below by the Compton wavelength, the extension of $\psi(\vec{k})$ in $k$-space is bounded from above by the corresponding wavenumber, and the extension of $\psi(\vec{p})$ in momentum-space is bounded by $Mc$. 

Since $\Delta |\vec{r}|$ and $\Delta |\vec{p}|$  are scalars, we may write these as $\Delta R_{\rm 3D}$ and $\Delta P_{\rm 3D}$, respectively, where $R_{\rm 3D} = |\vec{r}|$ and $P_{\rm 3D} = |\vec{p}|$. For approximately spherically symmetric wave packets, we expect 
\begin{equation} 
\langle{\hat{\vec{p}}}\rangle \simeq 0 \, , \ \ \ \Delta P_{\rm 3D} \simeq \sqrt{\langle{\hat{\vec{p}}^2}\rangle} \lesssim Mc \, , \label{R^2_sd}
\end{equation}
\begin{equation} 
\langle{\hat{\vec{r}}}\rangle \simeq 0 \, , \ \ \ \Delta R_{\rm 3D} \simeq \sqrt{\langle{\hat{\vec{r}}^2}\rangle} \gtrsim h/(Mc). \label{P^2_sd}
\end{equation}
The commutator of $\hat{\vec{r}}$ and $\hat{\vec{p}}$ is
\begin{eqnarray} \label{Comm_R^2P^2}
[\hat{\vec{r}},\hat{\vec{p}}] = i\hbar \, ,
\end{eqnarray} 
which implies 
\begin{eqnarray} \label{SUP_R^2P^2}
\Delta R_{\rm 3D} \, \Delta P_{\rm 3D} \geq \hbar/2 \, . 
\end{eqnarray} 
From Eqs.~(\ref{R^2_sd})-(\ref{P^2_sd}), it is therefore reasonable to make the identifications
\begin{equation} \label{P_ident}
\left(\Delta P_{\rm 3D}\right)_{\rm max} \simeq M c \, ,  \quad
\left(\Delta R_{\rm 3D}\right)_{\rm min} \simeq R_C \, , 
\end{equation}
where we will henceforth refer to the Compton wavelength as the Compton radius and restrict consideration to quasi-spherically symmetric distributions, the precise meaning of this term being explained in Sec.~\ref{Sec.5}. Under these conditions, the uncertainty relation for position and momentum allows us to recover the standard expression (\ref{ident}). 

The advantage of the above non-relativistic arguments 
is that they can be readily extended to the higher-dimensional case with extra compactified dimensions. The results obtained are phenomenologically robust, despite being derived in the approximate low-energy theory. 
However, the problem
 is that  the speed limit is put in by hand, without introducing additional relativistic effects, such as Lorentz invariance. 
So how does one extend the above argument to the relativistic case? Just as the non-relativistic relationship $E=p^2/(2M)$  corresponds to the Schr{\"o}dinger equation, so the relativistic relationship $E^2= M^2 c^4 + P^2c^2$ corresponds to the  Klein-Gordon equation, 
\begin{equation}
- \partial_t^2 \psi/c^2 + \nabla^2 \psi = (M c/ \hbar)^2  \psi \, .
\end{equation}
Looking for a plane-wave solution $e^{i( \vec{k} . \vec{r} - \omega t)}$ leads to the dispersion relation 
\begin{equation}
\omega^2 = c^2(k^2 - k_C^2)  \, ,
\end{equation}  
where $ k_C =Mc/\hbar$ is the reduced Compton wave-number. In the time-independent case, one has a spherically symmetric solution $\psi \propto e^{- k_C r}/r$, so 
the Compton wavelength can also be regarded as the scale on which the wave function decays or a correlation scale.

Another relativistic argument for the Compton wavelength is associated with pair-production. This combines the relativistic energy-momentum relation with the de Broglie relations (\ref{deBroglie2}). 
Since $E > Mc^2$ for $\lambda < R_{C}$, this shows that $R_C$ acts as a fundamental barrier beyond which pair-production occurs rather than further localization of the wave packet of the original particle. While the de Broglie wavelength marks the scale at which non-relativistic quantum effects become important and the classical concept of a particle gives way to the idea of a wave packet, the Compton wavelength marks the point at which relativistic quantum effects become significant and the concept of a single wave packet as a state wih fixed particle number  becomes invalid \cite{Berestetsky:1982aq}. $R_C$ is an effective minimum width because, on smaller scales, the concept of a single quantum mechanical particle breaks down and we must switch to a field description in which particle creation and annihilation occur in place of further spatial localization. 
However, as discussed in Appendix B, the minimum volume required for pair-production may be larger than $R_C^3$ 
when the wave packet is non-spherical.
This result is very relevant when we come to consider the higher dimensional case, especially in the context of particle production by black holes 

We have demonstrated that the existence of an effective cut-off for the maximum attainable energy/momentum in non-relativistic quantum mechanics implies the existence of a minimum attainable width for (almost) spherically symmetric wave functions, and this may be identified with the Compton radius for $P_{\rm 3D} \lesssim Mc$. For non-spherically symmetric systems we may still consider the 
upper bound on each momentum component, $p_i = \hbar k_i < Mc$, as giving rise to a 
lower bound for the spatial extent of the wave packet in $i^{th}$ spatial direction. However, as demonstrated in Appendix B, the minimum volume required for pair-production may be much larger than $R_C^3$. In the presence of compact extra dimensions, in which asymmetry is the norm, the existence of a {\it maximum} spatial extent (the compactification scale) also gives rise to a {\it minimum} momentum uncertainty. In particular, pair-production can occur for volumes exceeding $R_C^{3+n}$.  As we shall see in Sec. \ref{Sec.5}, this has important implications for the physics of quantum particles in compactified spacetimes, which have hitherto not been considered in the literature.

\section{Higher-dimensional black holes and TeV quantum gravity} \label{Sec.3}

The black hole boundary in Fig.~\ref{MR} assumes there are three spatial dimensions but many theories, including string theory, suggest that the dimensionality could increase on sufficiently small scales. Although the extra dimensions are often assumed to be compactified on the Planck length, there are also models \cite{rs,arkani,kanti2004} in which they are either infinite or compactified on a scale much larger than $R_P$, and these are the models of interest here. In this section, we will assume that the standard expression for the Compton wavelength applies even in the higher-dimensional case and explain why the existence of large extra dimensions could then lead to TeV quantum gravity and the production of black holes at accelerators. Although the argument is well-known, we present it in a way (cf. \cite{Ca:2013}) which is useful when we come to consider non-standard models.

For simplicity, we first assume that the extra dimensions are associated with a single length scale $R_E$. If the number of extra dimensions is $n$, then in the Newtonian approximation the gravitational potential generated by a mass $M$ is \cite{Maartens:2010ar,Maartens:2005ww}  
\begin{equation} \label{V1}
V_{\mathrm{grav}} =  \frac{G_D M}{R^{1+n}}  \quad (R < R_E) \, ,
\end{equation} 
where $G_D$ is the higher-dimensional gravitational constant and $D = 3+n+1$ is the number of spacetime dimensions in the relativistic theory. For $R>R_E$, the factor $R^{1+n}$ is replaced by $R R_E^n$, so one has 
\begin{equation} \label{V2}
V_{\mathrm{grav}} =  \frac{G M}{R}  \quad \mathrm{with}  \quad G = \left(\frac{G_D}{R_E^n}\right)   \quad (R > R_E)  \, .
 \end{equation}
Thus  one recovers the usual form of the potential in this region.
The higher-dimensional nature of the gravitational force is only manifest for $R < R_E$. This follows directly from the fact that general relativity can be extended to an arbitrary number of dimensions, so we may take the weak field limit of Einstein's field equations in $3+n+1$ dimensions for $R < R_E$. In the Newtonian limit, the effective gravitational constants at large and small scales are different because of the dilution effect of the extra dimensions.

When considering scenarios with many extra dimensions, dimensional analysis may cease to be reliable for numerical estimates. It works well with three dimensions because maximally symmetric volumes and areas scale as $V \sim R^3$ and $A \sim R^2$, respectively, with the numerical coefficients being of order unity. 
However, as pointed out by Barrow and Tipler \cite{bt}, the volume of an $n$-sphere of radius $R$ is $\pi^n (2R)^n \Gamma(1+n/2)$ in Euclidean space, so there is an extra numerical factor $(2\pi e/n)^{n/2} (n \pi)^{-1/2}$ which decreases exponentially for large $n$.
Similar deductions hold for (maximally symmetric) $n$-dimensional surface areas. Thus the dimensionless factors 
become important in higher dimensional spaces.

For present purposes, we may define an {\it effective} compactification scale $R_E \equiv \kappa(n)^{1/n}\mathcal{R}_E$, where $\mathcal{R}_E$ is the {\it true} compactification scale of the extra dimensions and $\kappa(n)$ is defined by the $n$-dimensional volume  being $V_{(n)} = \kappa(n) \mathcal{R}_E^{n}$. This yields $V_{(n)} \sim R_E^{n}$, as used in the estimate of $G$
 in Eq.~(\ref{V2}),
so the resulting expressions remain phenomenologically valid. 
Similar arguments can be used to define effective length scales corresponding to highly asymmetric distributions, the simplest 
being $\Delta R_{\rm 3D} \sim (\Delta x \Delta y \Delta z)^{1/3}$, for $\Delta x \neq \Delta y \neq \Delta z$, in three dimensions. As we will show 
 in Sec.~\ref{Sec.4}, in higher dimensions such {\it effective} characteristic length scales
quantify the asymmetry of a system and  play a key role in determining its physics.

There are two interesting mass scales associated with the length scale $R_E$: the mass whose Compton wavelength is $R_E$, 
\begin{eqnarray}  \label{M_E} 
M_{E} \equiv \frac{\hbar}{c R_E} \simeq  M_{P}\frac{R_{P}}{R_E} \, ,
\label{ME}
\end{eqnarray} 
and the mass whose Schwarzschild radius is $R_E$,
\begin{eqnarray}  \label{M_C} 
M_E' \equiv \frac{c^2 R_E}{G} \simeq M_{P} \frac{R_E}{R_{P}} \,  .
\label{ME'}
\end{eqnarray}  
These mass scales are reflections of each other in the line $M=M_{P}$ in Fig 1, so that $M_{E}' = M_{P}^2/M_{E}$. An important implication of Eq.~(\ref{V1}) is that the usual expression for the Schwarzschild radius no longer applies for masses below $M_E'$. If the black hole is assumed to be (approximately) spherically symmetric in the higher-dimensional space on scales $R \ll R_E$, the expression for $R_S$ must be replaced with
\begin{equation} \label{higherBH}
R_S \simeq R_E  \left(\frac{M}{M_E'} \right)^{1/(n+1)} \simeq R_{*}\left(\frac{M}{M_{P}}\right)^{1/(1+n)} \, ,
 \end{equation}
where $R_{*}$ 
 is defined by Eq.~(\ref{R_*}).
Therefore, the slope of the black hole boundary in Fig.~\ref{MR} becomes shallower for $M \lesssim M_E'$.

Strictly speaking, the metric associated with Eq.~(\ref{higherBH}) is only valid for infinite extra dimensions, since it assumes asymptotic flatness \cite{Hor12}. For black hole solutions  with compact extra dimensions, one must ensure periodic boundary conditions with respect to the compact space. However, Eq.~(\ref{higherBH}) should be accurate for black holes with $R_S \ll R_E$, so we adopt this for the entire range $R_{P} \lesssim R \lesssim R_E$ as a first approximation. Similar problems arise, even in the Newtonian limit, since Eq.~(\ref{V2}) is also only valid for infinite extra dimensions and does not respect the periodicity of the internal space. In practice, we expect corrections to smooth out the transition around $R_S \simeq R_E$, so that the true metric yields the asymptotic forms corresponding to the Schwarzschild radius of a $(3+1)$-dimensional black hole on scales $R_S \gg R_E$ and a $(3+n+1)$-dimensional black hole on scales $R_S \ll R_E$. 

This form of $R_S(M)$ for various values of $n$ is indicated in Fig.~2(a). The intersect with the Compton boundary (assuming this is unchanged) is then given by Eq.~(\ref{revisedplanck}). This implies $M_{P}'  \ll M_{P}$ and $R_{P}'  \gg R_{P}$ for $R_E  \gg R_{P}$. 
If the accessible energy is $E_{\rm max}$, then the  extra dimensions can only be probed for
\begin{equation}
R_E  > R_P \left( \frac{c^2M_P}{E_{\rm max}} \right)^{(2+n)/n} \simeq 10^{(32/n) - 17} \left(\frac{E_{\rm max}}{{\rm 10 TeV}}\right)^{-(2+n)/n}\mathrm{cm} \, ,
\end{equation}
where $E_{\rm max}$ is normalised to $10$~TeV, the order of magnitude energy associated with the Large Hadron Collider (LHC). Thus black holes can be created at the LHC providing
\begin{equation} \label{nconstraint}
R_E  > 10^{(30/n) - 18}\, \mathrm{cm}
\simeq 
\begin{cases}
10^{12}\, \mathrm{cm}
& (n=1) \\
10^{-3} \, \mathrm{cm} 
& (n=2) \\
10^{-14}\, \mathrm{cm} 
& (n=7) \\
10^{-18} \, \mathrm{cm} 
& (n=\infty) \, .
\end{cases}
\end{equation}
Clearly, $n=1$ is excluded on empirical grounds but $n=2$ is possible. One expects $n=7$ in M-theory \cite{Becker:2007zj}, so it is interesting that $R_E$ must be of order a Fermi if all the dimensions are large.
$R_E \rightarrow 10^{-18}$cm as $n \rightarrow \infty$ since this is the smallest scale which can be probed by the LHC. 

The above analysis assumes that all the extra dimensions have the same size. One could also consider a hierarchy of compactification scales, $R_i = \alpha_i R_{P}$ with $\alpha_1 \geq \alpha_2 \geq .... \geq \alpha_n \geq 1$, such that the dimensionality progressively increases as one goes to smaller distances \cite{Ca:2013}. In this case, the effective {\it average} length scale associated with the compact internal space is
\begin{equation} 
\langle R_E \rangle = \left( \prod_{i=1}^{n} R_i  \right)^{1/n} = R_{P} \left( \prod_{i=1}^{n} \alpha_i \right)^{1/n} \, .
\end{equation} 
and the new effective Planck scales are 
\begin{equation}
R_{P}' \simeq  \left(R_{P}^2 \prod_{i=1}^{n} R_i\right)^{1/(2+n)} \simeq (R_{P}^2 \langle R_E \rangle^n)^{1/(2+n)}  
\end{equation}
\begin{equation}
M_{P}' \simeq \left(M_{P}^2 \prod_{i=1}^{n}M_i^n\right)^{1/(2+n)} \simeq (M_{P}^2 \langle M_E \rangle ^{n})^{1/(2+n)} \, ,
\label{revisedplanck*}
\end{equation}
where $M_i \equiv \hbar/(cR_i)$ and $\langle M_E \rangle \simeq \hbar/(c \langle R_E \rangle )$. For $R_{k+1} \lesssim R \lesssim R_{k}$, the effective Schwarzschild radius is then given by 
\begin{equation} \label{effectiveschwarz}
R_S = R_{*(k)}\left(\frac{M}{M_{P}}\right)^{1/(1+k)}, \ \ \ R_{*(k)} = \left(R_{P} \prod_{i=1}^{k \leq n}R_{i}\right)^{1/(1+k)}.
\end{equation}
This situation is represented in Fig.~2(b). Clearly, for given $n$, the Planck scales are not changed as much
as in the scenario for which the extra dimensions all have the same scale. 

\begin{figure}  \label{MR2}
\begin{center}
   \psfig{file=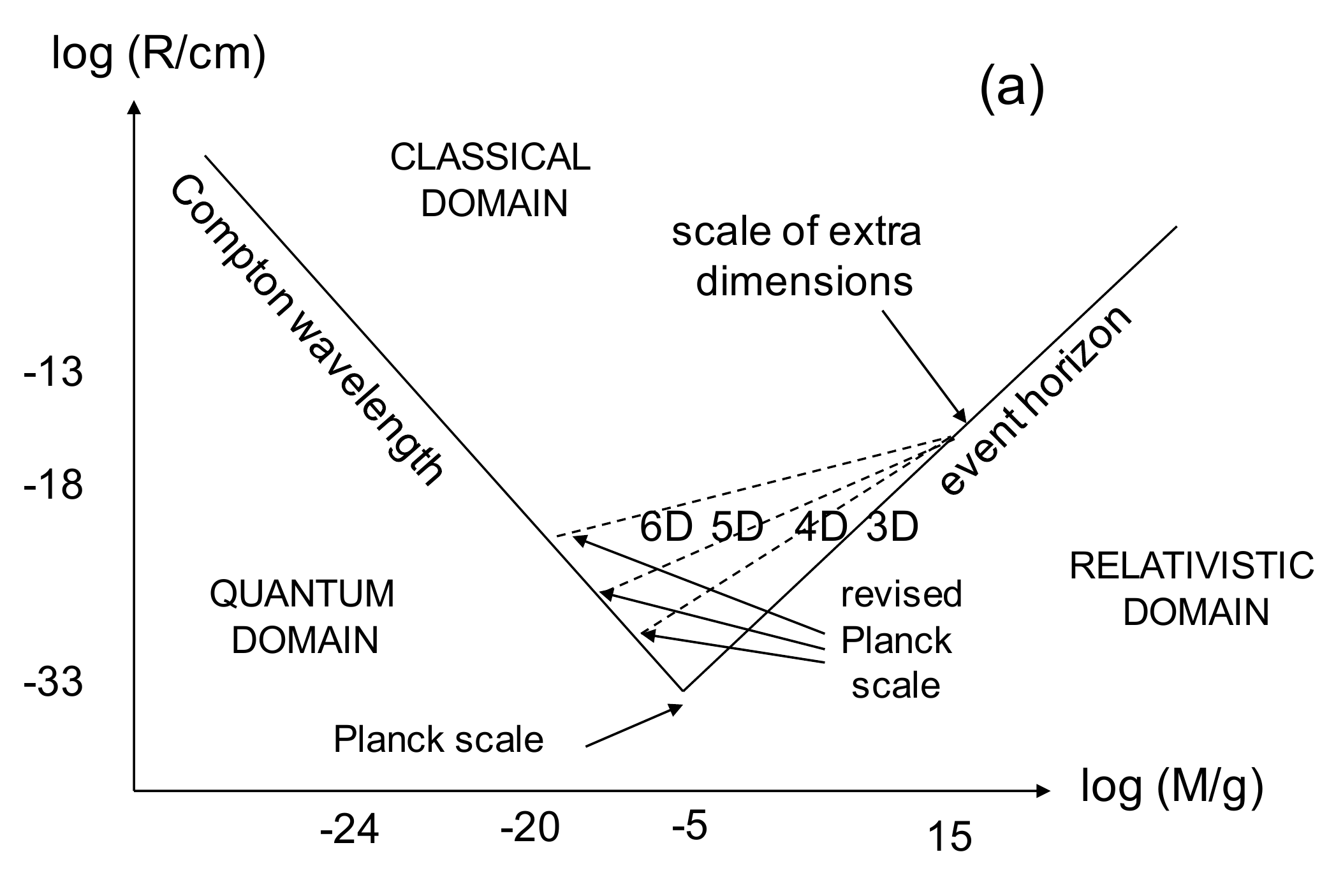,width=3.2in}
   \psfig{file=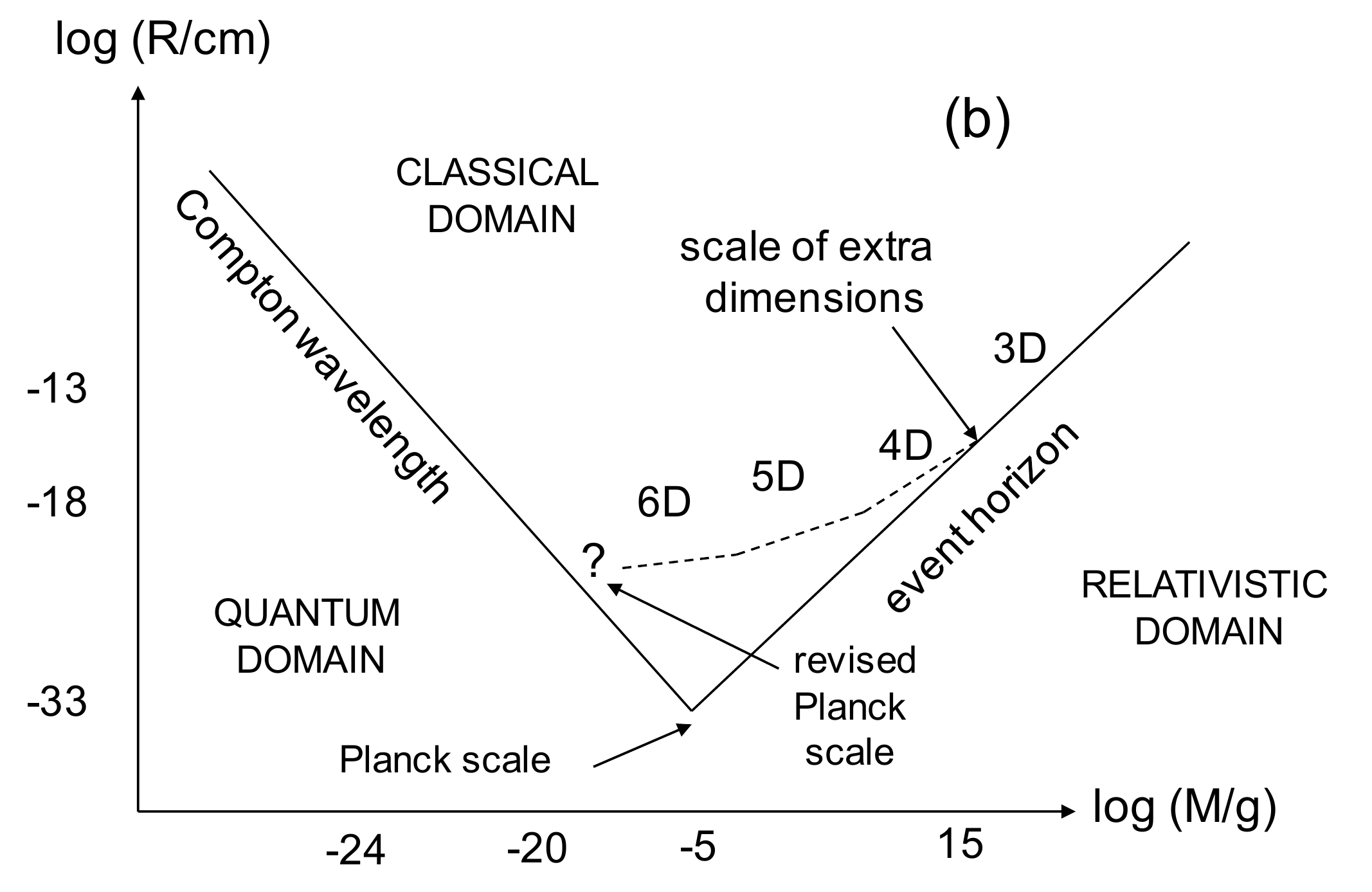,width=3.2in}
  \caption {Modification of the Schwarzschild line in the $(M,R)$ diagram in the presence of extra compact dimensions associated with a single length scale (a) or a hierarchy of length scales (b). If the Compton scale preserves its usual form, the effective Planck scales are shifted as indicated.} 
\end{center}
\end{figure} 

\begin{figure}  \label{Fig3}
\begin{center}
   \psfig{file=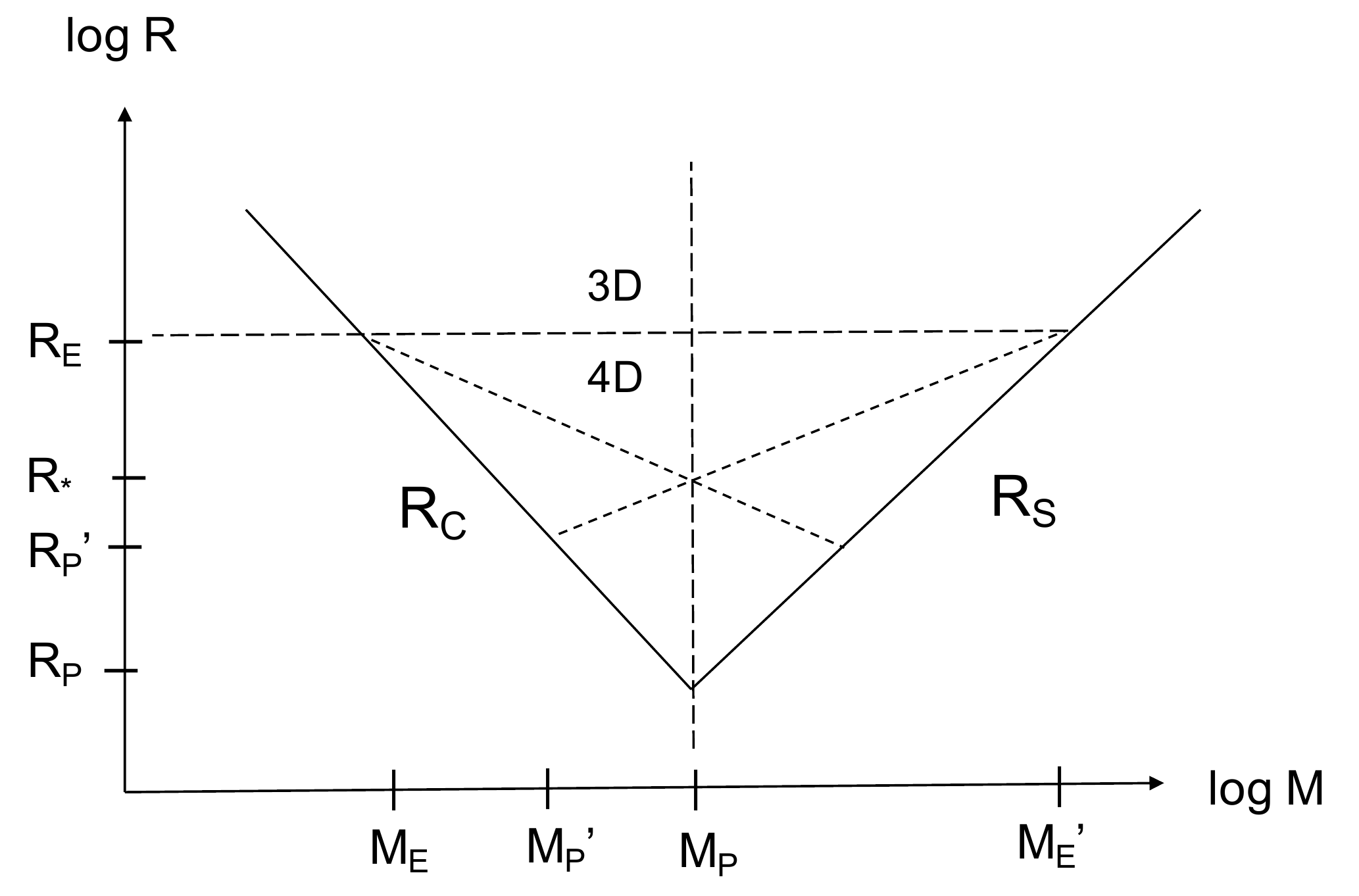,width=3.0in}
  \caption {Key mass and length scales in the 3D case (solid lines) and 4D cases (dotted lines) if the extra dimension is compactified on a scale $R_E$. The associated Compton and Schwarzschild masses are $M_E$ and $M_E'$, respectively. The revised Planck scales are $M_{P}'$ and $R_{P}'$ if duality is violated but $M_{P}$ and $R_*$ if it is preserved.}  
\end{center}
\end{figure}

The relationship between the various key scales ($R_E, R_E',R_{P},R_{P}',M_{P},M_{P}',R_*$) in the above analysis is illustrated in Fig.~\ref{Fig3} for the case of one extra spatial dimension ($n=1$). This shows that the duality between the Compton and Schwarzschild length scales is lost  if one introduces extra spatial dimensions. However, this raises the issue of whether the expression for the Compton wavelength should also be modified in the higher-dimensional case. We argue below that in this scenario a phenomenologically important length scale is the {\it effective} Compton wavelength, which may be identified with the minimum effective width (in $3$-dimensional space) of the higher-dimensional wave packet $(\Delta x)_{\rm min}$.

\section{Uncertainty Principle in higher dimensions} \label{Sec.4}

In this section, we consider whether the uncertainty in the momentum $\Delta P$ in a ($3+n+1$)-dimensional spacetime with $n$ compact dimensions scales inversely with the uncertainty in the position $\Delta R$, as in the $3$-dimensional case, or according to a different law. If we interpret $\Delta R$ to mean  the localisability of a particle, in the sense discussed in Sec.~\ref{Sec.2}, we find that this depends crucially on the 
distribution of the wave packet in the extra dimensions, i.e. on the degree of asymmetry between its size in the infinite and compact dimensions.i

In 3-dimensional space with Cartesian coordinates $(x,y,z)$, the uncertainty relations for position and momentum are
\begin{equation} 
\Delta x\Delta p_{x} \gtrsim \hbar \, , \quad
\Delta y\Delta p_{y} \gtrsim \hbar \, , \quad
\Delta z\Delta p_{z} \gtrsim \hbar \label{3D_URz} \, .
\end{equation}
For spherically symmetric distributions, we have
\begin{eqnarray}  \label{SpherSymm3D-1}
\Delta x \simeq \Delta y \simeq \Delta z \simeq \Delta R_{\rm 3D} \, , \quad
\Delta p_{x} \simeq \Delta p_{y} \simeq \Delta p_{z} \simeq \Delta P_{\rm 3D} \, ,
\end{eqnarray}
where the axes are arbitrarily orientated, so that the relations (\ref{3D_URz}) are each equivalent to
\begin{eqnarray}  \label{3D_UR_SpherSymm} 
\Delta R_{\rm 3D} \, \Delta P_{\rm 3D} \gtrsim c R_{P}M_{P} = \hbar \, .
\end{eqnarray}
In $(3+n)$ spatial dimensions, we also have
\begin{eqnarray} \label{(3+n)-D_UR}
\Delta x_i \, \Delta p_{i} \gtrsim \hbar \quad (i = 1,2, . . . n) \, ,
\end{eqnarray}
so for distributions that are spherically symmetric with respect to the three large dimensions we obtain 
\begin{eqnarray}  \label{(3+n)-D_UR_combined} 
\Delta R_{\rm 3D} \, \Delta P_{\rm 3D} \left(\prod_{i=1}^{n} \Delta \, x_i\Delta p_{i}\right)  \gtrsim  \hbar^{1+n} \, ,
\end{eqnarray}
where $\Delta \, x_i \neq \Delta R_{\rm 3D}$ and $\Delta p_{i} \neq \Delta P_{\rm 3D}$ generally. The exponent on the right is $1+n$, rather than $3+n$, because there is only {\it one} independent relation associated with the large spatial dimensions due to spherical symmetry. Assuming, for simplicity, that the extra dimensions are compactified on a single length scale $R_E$, then totally spherically symmetric wave functions are only possible on scales $\Delta R < R_E$ in position space or $\Delta P > c M_E$ in momentum space. In this case, we may identify the standard deviations in the extra dimensions (i.e. in both position and momentum space) with those in the infinite dimensions, 
\begin{eqnarray}  \label{SpherSymm(3+n)D-1} 
\Delta x_{i} \simeq \Delta R_{\rm 3D} \, , \quad
\Delta p_{i} \simeq \Delta P_{\rm 3D} \, , 
\end{eqnarray}
for all $i$, so that Eq.~(\ref{(3+n)-D_UR_combined}) reduces to (\ref{3D_UR_SpherSymm}). Following the usual identifications, this gives the standard expression for the Compton wavelength in a higher-dimensional context. 

However, this is not the only possibility. The condition of spherical symmetry in the three large dimensions implies that the directly observable part of $\psi$ is characterized by a single length scale, the 3-dimensional radius of the wave packet $\Delta R_{\rm 3D}$. One can therefore characterise the physical distribution of the wave packect by the ($1+n$)-dimensional volume 
\begin{eqnarray} \label{Volume}
V_{(1+n)} \simeq \Delta R_{\rm 3D} \prod_{i=1}^{n}\Delta x_i \equiv (\Delta \mathcal{R})^{1+n} \, , 
\end{eqnarray}
where $\Delta \mathcal{R}$ corresponds to the effective ($1+n$)-dimensional radius of the particle.
As demonstrated in Appendix B, for wave packets that are spherically symmetric in the large directions but irregular in the compact space, it is this length scale 
 which controls pair-production
rather than the geometric average over all $3+n$ dimensions, $[(\Delta R_{\rm 3D})^3\Pi_{i=1}^{n} \Delta x_i]^{1/(3+n)}$.
Indeed, Appendix B suggests that 
$\Delta \mathcal{R}$ is the key length scale if only independent uncertainty relations contribute to the composite measurement. This makes sense, since were we able to isolate our measurements of the 3-dimensional part of the wave packet, this would yield only a single length scale $\Delta R_{\rm 3D}$; any ``smearing" of this measurement due to the spread of the wave packet in the extra dimensions must be due to the $n$ additional {\it independent} widths, $\Delta x_i \neq \Delta R_{\rm 3D}$. 

We now consider scenarios with $\Delta R_{\rm 3D} \neq \Delta \mathcal{R}$ and $\Delta x_i \neq \Delta \mathcal{R}$, for at least some $i$. Such states may be considered ``quasi-spherical" in the sense that they are spherically symmetric with respect to the three large dimensions but 
possibly extremely irregular from the higher-dimensional perspective. 
Note that  $\Delta R_{\rm 3D} < R_{E}$ for $M>M_{E}$ 
and in this case
Eq.~(\ref{(3+n)-D_UR_combined}) becomes 
\begin{eqnarray} \label{(3+n)-D_UR*1}
(\Delta \mathcal{R})^{1+n}\Delta P_{\rm 3D} \left(\prod_{i=1}^{n} \Delta p_{i}\right) \gtrsim  \hbar^{1+n} \, .
\label{volbound}
\end{eqnarray}
We restrict ourselves to states for which
\begin{eqnarray}  \label{SpherSymm(3+n)D-2**} 
\Delta p_{i} \simeq \kappa_{i}^{-1} cM_{P} \, , 
\end{eqnarray} 
where the $\kappa_i$ are dimensionless constants satisfying
\begin{eqnarray} \label{X}
1 \leq \kappa_i \leq \frac{R_E}{R_{P}} = \frac{M_{P}}{M_E} \, .
\end{eqnarray} 
This ensures that 
\begin{eqnarray} \label{P_range}
cM_{P} \geq \Delta p_i \geq cM_E
\end{eqnarray} 
and restricts us to the higher-dimensional region of the $(M,R)$ diagram. Conditions (\ref{(3+n)-D_UR}) then reduce to
\begin{eqnarray} \label{(3+n)-D_UR*}
\Delta x_i \gtrsim \kappa_i R_{P} \, ,
\end{eqnarray}
which together with Eq. (\ref{X}) ensures
\begin{eqnarray}
R_{P} \leq (\Delta x_i)_{\rm min} \leq R_E \, .
\end{eqnarray} 
Note that we are still considering the case in which all extra dimensions are compactified on a single length scale $R_E$, so the $\kappa_i$ have no intrinsic relationship with the constants $\alpha_i$ used to characterize the hierarchy of length scales in Sec.~\ref{Sec.3}. They  simply paramaterize the degree to which each extra dimension is ``filled" by the wave packet (e.g. if $\kappa_i = 1$, the physical spread of the wave packet in the $i^{th}$ extra dimension is $R_{P}$). Were we to consider a similar parameterization in the hierarchical case, it would follow immediately that $\kappa_i \leq \alpha_i$.

Equation (\ref{(3+n)-D_UR*1}) now becomes
\begin{eqnarray}  \label{(3+n)-D_UR_combined**} 
\Delta \mathcal{R} \gtrsim R_{P}\left[\frac{cM_{P}(\prod_{i=1}^{n}\kappa_{i})}{\Delta P_{\rm 3D}}\right]^{1/(1+n)} \, .
\end{eqnarray}
The validity of this bound is subject to the quasi-spherical symmetry condition (\ref{Volume}) but it is \emph{stronger} than the equivalent condition (\ref{3D_UR_SpherSymm}) for fully spherically symmetric states
(i.e. the lower limit on $\Delta \mathcal{R}$ falls off more slowly with increasing  $\Delta P_{\rm 3D}$). 
By definition, such a wave packet is also quasi-spherically symmetric in momentum space, in the sense that it is spherically symmetric with respect to three infinite momentum dimensions, but not with respect to the full $(3+n)$-dimensional momentum space. 

For fully spherically symmetric states, we must put each $\kappa_i$ equal to a single value $\kappa$, so that $\Delta \mathcal{R} \equiv \Delta R_{\rm 3D}\simeq \kappa R_{P}$ and $\Delta P_{\rm 3D} \simeq \kappa^{-1} cM_{P}$, since the momentum space representation $\psi(P)$ is given by the Fourier transform of $\psi(R)$. Hence a wave function that is totally spherically symmetric in the $3+n$ dimensions of position space will also be totally spherically symmetric in the $3+n$ dimensions of momentum space. 
For quasi-spherical states, 
the volume occupied by the particle in the $n$ extra dimensions of momentum space is
\begin{eqnarray} \label{Volume*}
V_{p(n)} \simeq \prod_{i=1}^{n}\Delta p_i \simeq (cM_{P})^{n}\left(\prod_{i=1}^{n}\kappa_{i}\right)^{-1}.
\end{eqnarray}
In these states, we will assume that the extra-dimensional momentum volume remains {\it fixed} but the total momentum volume also depends on the 3-dimensional part $\Delta P_{\rm 3D}$, which may take any value satisfying Eq.~(\ref{(3+n)-D_UR_combined**}). We also fix the extra-dimensional physical volume.

The underlying physical assumption behind the mathematical requirement of fixed extra-dimensional volume is 
that the extra-dimensional space can only be probed indirectly -- for example, via high-energy collisions between particles whose momenta in the compact directions cannot be directly controlled. Therefore the net effect of any interaction is likely to leave the total extra-dimensional volume occupied by the wave packet unchanged, even if its 3-dimensional part can be successfully localized on scales below $R_E$. This is the mathematical expression of the fact that we have no control over the extra-dimensional part of {\it any} object - including that of the apparatus used to probe the higher-dimensional system. As such, complete spherical symmetry in the higher-dimensional space is not expected
and the most natural assumption is that asymmetry persists between the 3-dimensional and extra-dimensional parts of the wave function.  Indeed, the most natural assumption is $\Delta x_i = R_i = R_E$ 

Since Eq.~(\ref{X}) implies
\begin{eqnarray}  \label{kappa_choice}
1 \leq \prod_{i=1}^{n}\kappa_{i} \leq \left(\frac{R_{E}}{R_{P}}\right)^{n},
\end{eqnarray}
we have
\begin{eqnarray}  \label{(3+n)-D_UR_combined***} 
R_{P}\left(\frac{cM_{P}}{\Delta P_{\rm 3D}}\right)^{1/(1+n)} \lesssim (\Delta \mathcal{R})_{\rm min} \lesssim R_{*}\left(\frac{cM_{P}}{\Delta P_{\rm 3D}}\right)^{1/(1+n)},
\end{eqnarray}
where $R_{*}$ is defined by Eq. (\ref{R_*}) and we restrict ourselves to the higher-dimensional region of the $(M,R)$ diagram,
\begin{eqnarray} \label{C}
cM_{P} \geq \Delta P_{\rm 3D} \geq cM_E \, . 
\end{eqnarray}
In the extreme case $\Delta P_{\rm 3D} = cM_{P}$, this gives
\begin{eqnarray} \label{B}
R_{P} \leq (\Delta \mathcal{R})_{\rm min} \leq R_*
\end{eqnarray}
for \emph{any} choice of the constants $\kappa_i$, with the extreme limits $(\Delta \mathcal{R})_{\rm min} = R_{P}$ and $(\Delta \mathcal{R})_{\rm min} = R_*$ corresponding to $\kappa_i \rightarrow 1$ and $\kappa_i \rightarrow R_E/R_{P}$, respectively. 

The first limit corresponds to the scenario $R_E \rightarrow R_* \rightarrow R_{P}$, which recovers the standard Planck length bound on the minimum radius of a Planck mass particle.
In other words, if {\it all} the extra dimensions are compactified on the Planck scale, both the standard $3$-dimensional Compton and Schwarzschild formulae hold all the way down to $R_{P}$, giving the familiar intersect. However, if $R_E > R_{P}$, then $(\Delta \mathcal{R})_{\rm min}$ for a Planck mass particle is larger than the Planck length and may be as large as the critical value $R_*$. This is the second limit and it occurs when the higher-dimensional part of the wave packet completely ``fills" the extra dimensions, 
each of these being compacified on the scale $R_E$.

For $\Delta P_{\rm 3D} \leq cM_E$, the same scenario gives $(\Delta \mathcal{R})_{\rm min} \geq R_E$, which corresponds to the effectively 3-dimensional region of the $(M,R)$ plot. In this region, the assumption of quasi-sphericity breaks down and the 3-dimensional and higher-dimensional parts of the wave packet decouple with respect to measurements which are unable to probe the length/mass scales associated with the extra dimensions. We may therefore set
\begin{eqnarray}  \label{(3+n)-D_UR_combined***A} 
\Delta \mathcal{R} \gtrsim R_{*}\left(\frac{cM_{P}}{\Delta P_{\rm 3D}}\right)^{1/(1+n)}
\end{eqnarray}
as the strongest \emph{lower} bound on $\Delta \mathcal{R}$, since this is the \emph{upper} bound on the value of $(\Delta \mathcal{R})_{\rm min}$. To reiterate, this comes from combining two assumptions: (a) the wave function of the particle is quasi-spherically symmetric -- in the sense of  Eq.~(\ref{Volume}) -- with respect to the full higher-dimensional space on scales $\Delta \mathcal{R} \lesssim R_E$; and (b) the wave packet is space-filling in the $n$ additional dimensions of position space.

Note that the unique 3-dimensional uncertainty relation and each of the $n$ independent uncertainty relations for the compact directions still hold individually. However, the higher-dimensional uncertainty relations are satisfied for any choice of the constants $\kappa_i$ in the range specified by Eq.~(\ref{X}) and the remaining 3-dimensional relation $\Delta R_{\rm 3D} \gtrsim \hbar/\Delta P_{\rm 3D}$ is satisfied automatically for any $\Delta P_{\rm 3D}$ satisfying Eq.~(\ref{(3+n)-D_UR_combined**}). In the limit $\kappa_i \rightarrow R_E/R_{P}$ for all $i$, in which the wave packet completely fills the compact space, 
we have $(\Delta R_{\rm 3D})_{\rm min} = (\Delta \mathcal{R})_{\rm min} = R_E$ when $\Delta P_{\rm 3D} = cM_E$, so that the 3-dimensional and $(3+n)$-dimensional formulae match seamlessly. Thus, for $\Delta P_{\rm 3D} \simeq cM_{P}$, we have $(\Delta \mathcal{R})_{\rm min} \simeq R_*$ but the genuine 3-dimensional radius of the wave packet is of order $(\Delta R_{\rm 3D})_{\rm min} \simeq R_{P}$. 

\section{Compton wavelength and black holes in higher dimensions} \label{Sec.5}

In discussing the Compton wavelength of a particle in higher dimensions, the question of the experimental accessibility of the extra dimensions is crucial. Unless the experimental set-up allows direct control over the size of the wave-packet in the compact space, we cannot assume that a probe energy $E \sim \hbar c/R_C$ implies the localisation of the (3+n)-dimensional wave-function within a volume $V \sim R_C^{3+n}$. If there is only control over the three large dimensions, then the total volume of the wave-packet may be much larger than the minimum value. In principle, this corresponds to a larger minimum width for the wave-packet (i.e. a larger effective higher-dimensional Compton wavelength).

In this section we use the analysis of Sec.~4 and the identifications (\ref{P_ident}) to derive an expression for the {\it effective} higher-dimensional Compton wavelength:
\begin{eqnarray} \label{HD_Compton+} 
R_{C} \simeq R_{E}\left(\frac{M_{E}}{M}\right)^{1/(1+n)} \simeq R_{*}\left(\frac{M_{P}}{M}\right)^{1/(1+n)} \, .
\end{eqnarray}
This is equivalent to the identifications
\begin{eqnarray}  \label{HD_Compton+*} 
R_C \simeq \Delta \mathcal{R} \, , \quad \Delta P_{3D} \simeq M c \, ,
\end{eqnarray}
where $ \Delta \mathcal{R}$ is given by Eq. (\ref{(3+n)-D_UR*1})
with $\Delta p_i \simeq M_Ec$ for all $i$. Clearly, Eq. (\ref{HD_Compton+}) is consistent with
the bound (\ref{volbound})
since $M<M_P$ in the particle regime. These arguments imply that, when extrapolating the usual arguments for the Compton wavelength in non-relativistic quantum theory to the case of compact extra dimensions, we should identify the geometric average of the spread of the wave packet in $1+n$ spatial dimensions with the effective particle `radius' $ \Delta \mathcal{R}$ but its spread in the large dimensions of momentum space with the rest mass. 

The identifications (\ref{HD_Compton+})-(\ref{HD_Compton+*}) are also consistent with the relativistic interpretation of the Compton wavelength as the minimum localization scale for the wave packet below which pair-production occurs (see Appendix B). 
This is fortunate, as there is clearly a problem with identifying the standard deviation of the \emph{total} higher-dimensional momentum, $P_T = \sqrt{P_{\rm 3D}^2 + P_{E}^2}$ where $P_{E}^2 = \Sigma_{i=1}^{n}p_ip^i$, with the rest mass of the particle. Since the standard deviations of the individual extra-dimensional momenta are bound from below by $\Delta p_i \geq cM_E$, we have  $\Delta P_T \geq c M_E$. 
The identification $\Delta P_T = M c$ would then imply $M > M_E$. Since $R_E$ must be very small to have avoided direct detection, $M_E$ must be large and the above requirement contradicts known physics as it requires \emph{all} particles to have masses $M > M_E$. 

The manifest asymmetry of the wave packet in position space on scales less than $R_E$ (and on scales greater than $cM_E$ in momentum space) also requires identifications of the form~(\ref{P_ident}) in order for the standard Compton formula to hold for $R \geq R_E$ in a higher-dimensional setting. What happens to the standard formula below this scale is unclear. If the wave packet is able to adopt a genuinely spherically symmetric configuration in the full higher-dimensional space (including momentum space), then the above arguments suggest the identifications $(\Delta R_{\rm T})_{\rm min} \simeq R_C$ and $(\Delta P_T)_{\rm max} \simeq Mc$ for $M_E \leq M \leq M_{P}$, so that the usual Compton formula holds all the way down to $M \simeq M_{P}$. The possible short-comings of this approach are that it would be valid only for spherically symmetric states and that it requires a change in the identification of the rest mass and particle radius at $M = M_E$, i.e. $R_C \simeq (\Delta R_{\rm T})_{\rm min} \rightarrow R_C \simeq (\Delta R_{\rm 3D})_{\rm min}$ and $Mc \simeq (\Delta P)_{\rm max} \rightarrow Mc \simeq (\Delta P_{\rm 3D})_{\rm max}$.

As wave packets will generally be asymmetric on scales $R \geq R_E$, it is reasonable to assume that the asymmetry will persist, even when we are able to (indirectly) probe 
scales associated with the extra dimensions. 
For example, we may consider the following two gendanken experiments.\\

\noindent (i) 
We localize a particle in 3-dimensional space by constructing a spherically symmetric potential barrier. We then gradually increase the steepness of the potential well, increasing the energy and localizing the particle on ever smaller length scales. In principle, we may even shrink the 3-dimensional radius below the scale of the internal space. But what about the width of the wave packet in the compact directions? Since we did not design our initial potential to be spherically symmetric in $3+n$ spatial dimensions -- having no direct manipulative control over its form in the extra dimensions -- it is unlikely that one would suddenly obtain a fully spherically symmetric potential in higher-dimensional space simply by increasing the energy at which our ``measuring device" operates (i.e. above $M_E c^2$).\\

\noindent (ii) 
We confine a particle within a spherical
region of 3-dimensional space by bombarding it with photons from multiple angles. Increasing the energy of the photons then reduces the radius of the sphere. But how can we control the trajectories of the probing photons in the internal space? Since, again, we do not have direct manipulative control  in the compact space over the apparatus that creates the photons, it is impossible to ensure anything other than a random influx of photons (with random extra-dimensional momenta) in the $n$ compact directions. In this case, we would expect to be able to measure the average photon energy and to relate this to a single \emph{average} length scale, but we cannot ensure exact spherical symmetry with respect to all $3+n$ dimensions, or measure the spread of the wave packet in each individual extra dimension.
In the most extreme case, we may expect the combined effects of our (random) experimental probing of the extra dimensions to cancel each other out, leaving the total volume of the wave packet in the compact space unchanged. This justifies Eq.~(\ref{Volume*}) but does not alter the reduction of the 3-dimensional
and hence {\it overall} volume of the wave packet when the energy of the probe particles/potential barrier is increased.\\

\noindent Together, these considerations lead to the scaling predicted by Eq.~(\ref{HD_Compton+}). As this corresponds to the maximum possible asymmetry for which a single length scale can be associated with $\psi$, this should give the highest possible lower bound on the size of a quantum mechanical particle in a spacetime with $n$ compact extra dimensions. 

Since the particle and black hole regimes are connected, it is also meaningful to ask if we can associate a wave function $\psi$ with a black hole. If so, should $\psi$ be associated with the centre of mass of the black hole or with its event horizon at $R_S<R_E$ (cf. Casadio \cite{casadio})? For classical non-extended bodies, i.e. point-particles, such problems of quantization do not arise. In the classical theory, with only infinite dimensions, a Schwarzschild black hole is the unique spherically symmetric vacuum solution \cite{Bir23}. However, in the quantum mechanical case, our previous analysis suggests that it may be possible to associate multiple quasi-spherically symmetric wave packets with the unique classical solution, just as we can for classical (spherically symmetric) point particles. The investigation of both these points lies beyond the scope of this paper and is left for future work.

To summarize our results for  higher-dimensional black holes and fundamental particles, we have 
\begin{equation}  \label{HD_Compton*} 
R_{C} \simeq 
\begin{cases}
R_{P}\frac{M_{P}}{M}
& (R_{C} \gtrsim R_E) \\
R_{*}\left(\frac{M_{P}}{M}\right)^{1/(1+n)} 
& (R_{C} \lesssim R_E)
\end{cases}
\end{equation}
\begin{equation}  \label{HD_Schwarz} 
R_{S} \simeq 
\begin{cases}
R_{P}\frac{M}{M_{P}} 
& (R_{S} \gtrsim R_E) \\
R_{*}\left(\frac{M}{M_{P}}\right)^{1/(1+n)} 
&(R_{S} \lesssim R_E)
\end{cases}
\end{equation}
for $n$ extra dimensions compactified on a single length scale $R_E$, and these lines intersect at $(R_{*},M_{P})$.
The crucial point is that there is no TeV quantum gravity in this scenario since the intersect of the Compton and Schwarzschild lines still occurs at $M \simeq M_{P}$. The effective Planck length is 
increased to $R_*$ but this does not allow the production of higher-dimensional black holes at accelerators. Thus, the constraint (\ref{nconstraint}) on the scale $R_E$ in the conventional picture no longer applies. 
\begin{figure}  \label{MR3}
\begin{center}
   \psfig{file=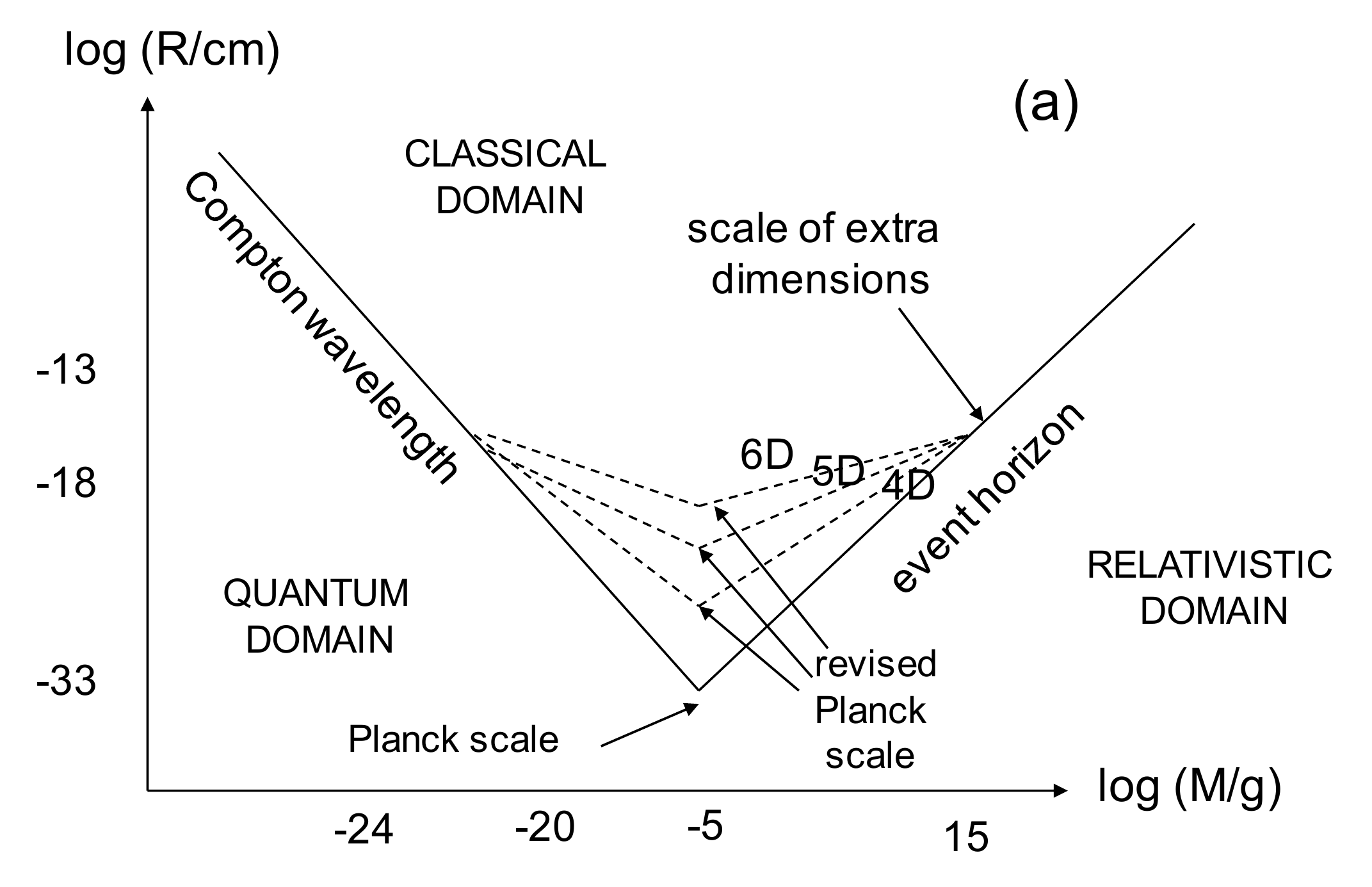,width=3.2in}
   \psfig{file=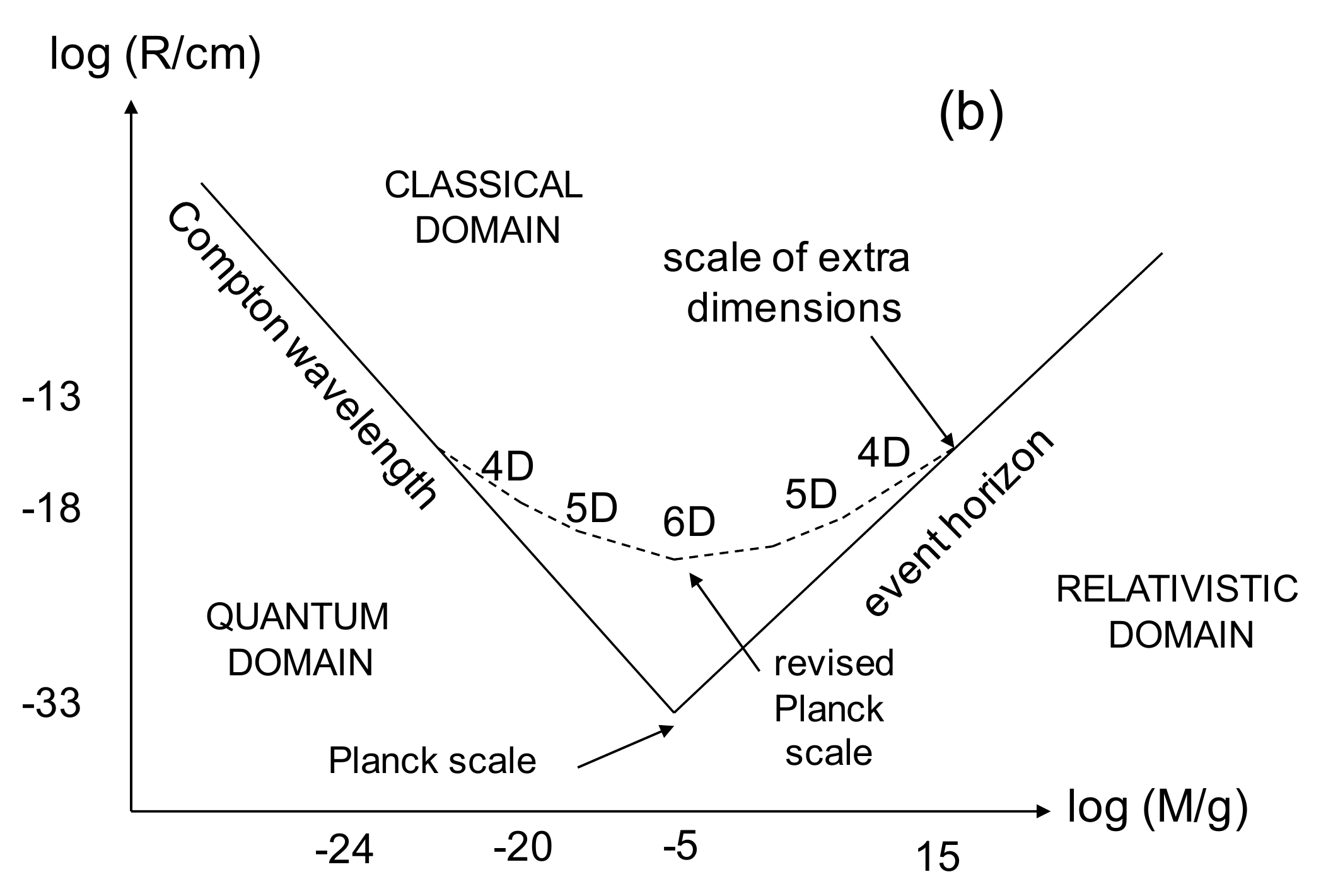,width=3.2in}
   \caption {Modifications of Fig.~2 for extra dimensions compactified on a single length scale (a) or a hierarchy of length scales (b) if one imposes quasi-spherical symmetry on the higher-dimensional wave packet, preserving the duality between the Compton and Schwarzschild expressions.}
\end{center}
\end{figure}

This scenario is illustrated in Fig.~4(a) for extra dimensions compactified on a single length scale $R_E$ and in Fig.~4(b) for a hierarchy of length scales. In the latter case, the expressions (\ref{HD_Compton*})-(\ref{HD_Schwarz}) must be modified to
\begin{equation} 
R_{C} \simeq 
\begin{cases}
R_{P}\frac{M_{P}}{M} 
& (R_{C} \gtrsim R_1) \\
R_{*(k)}\left(\frac{M_{P}}{M}\right)^{1/(1+k)} 
& (R_{k+1} \lesssim R_{C} \lesssim R_{k})
\end{cases}
\end{equation}
\begin{equation}  
R_{S} \simeq
 \begin{cases}
R_{P}\frac{M}{M_{P}} 
&(R_{C} \gtrsim R_1) \\
R_{*(k)}\left(\frac{M}{M_{P}}\right)^{1/(1+k)} 
& (R_{k+1} \lesssim R_{C} \lesssim R_{k})
\end{cases}
\end{equation}
where $R_1$ is the largest compact dimension and $R_{*(k)}$ is defined in Eq. (\ref{effectiveschwarz}). 

\section{Observational consequences} \label{Sec.6}

In this section we consider two possible observational consequences of retaining the duality between the Compton and Schwarzschild expressions. The first relates to the detectability of exploding primordial black holes (PBHs). There is still no unambiguous detection of such explosions but it has been claimed that some short-period gamma-ray bursts  could be attributed to PBHs \cite{cline}. The second relates to high-energy scattering experments and the enhancement of pair-production at at accelerators on scales below $R_E$. 

\subsection{Black hole evaporation}  \label{Sec.7.1}

The Hawking temperature of a black hole of mass $M$ and radius $R_S$ in three dimensions is 
\begin{eqnarray}  \label{T_H} 
T_H \simeq T_{P}\frac{M_{P}}{M} \simeq T_{P} \frac{R_{P}}{R_S} \, ,
\end{eqnarray}
where $T_P = M_Pc^2/k_B$ is the Planck temperature. This result can be obtained from the relations
\begin{eqnarray} 
\Delta R_{\rm 3D} \simeq R_S \propto M , \quad \Delta P_{\rm 3D} \propto 1/(\Delta R_{\rm 3D}), \quad T_H \propto \Delta P_{\rm 3D} \propto  M^{-1}  \, ,
\label{3D argument}
\end{eqnarray} 
where the second relation is the standard Uncertainty Principle and the third relation assumes a black-body distribution for the emitted particles. The temperature can also be obtained from the surface gravity:
\begin{equation}
T_H \propto \kappa  \propto M/R_S^{2} \propto M^{-1} \, ,
\end{equation}
this being equivalent  to the relations
\begin{eqnarray}
  \label{T_H_ident} 
\Delta R_{\rm 3D} \simeq R_S \propto M , \quad \Delta P_{\rm 3D} \propto 1/(\Delta R_{\rm 3D}), \quad T_H \propto 1/ \Delta R_{\rm 3D} \propto  M^{-1}  \, .
\end{eqnarray} 
In this formulation the second expression is not needed to derive $T_H$ but is required by the HUP. The only difference between (\ref{3D argument}) and (\ref{T_H_ident}) is that the first associates the temperature with a momentum and the second with a length but both  sets of identifications yield Eq.~(\ref{T_H}).

In the higher-dimensional case, if all the extra dimensions have the same compactification scale $R_E$ and one assumes the standard expression for the Compton wavelength, then the temperature is modified to \cite{CaDaMa03,CaDa04} 
\begin{eqnarray}  \label{T_H_HD} 
T_H \simeq 
T_P' \left(\frac{M_{P}'}{M}\right)^{1/(1+n)} \simeq T_*\left(\frac{M_{P}}{M}\right)^{1/(1+n)} .
\end{eqnarray}
Here $M_P'$ is given by Eq. (\ref{revisedplanck}) and 
we have used the definitions
\begin{eqnarray}  \label{T*} 
T_P' \equiv M_P' c^2/k_B, \quad  T_*  \equiv (T_PT_E^n)^{1/(1+n)}, \quad T_E \equiv M_Ec^2/k_B  \simeq T_{P} R_{P}/R_E \, .
\end{eqnarray}
The $M$-dependence in Eq.~(\ref{T_H_HD}) can be derived from the relations
\begin{eqnarray} 
\ \Delta R_{\rm T} \simeq R_S \propto M^{1/(1+n)} , \quad \Delta P_T \propto 1/(\Delta R_{\rm T}) , \quad T_H \propto \Delta P_T \propto  M^{-1/(1+n)} \, ,
\end{eqnarray} 
where $\Delta P_T$ and $\Delta R_T$ appear because the wave function is assumed to be spherically symmetric in the full space.
The temperature can again be obtained from the surface gravity:
\begin{equation}
T_H \propto  \kappa \propto M/R_S^{2+n} \propto M^{-1/(1+n)} \, .
\label{HDSG}
\end{equation}
Note that Eq.~(\ref{T_H_HD}) extends all way down to the reduced Planck scale $M_P'$, where the temperature has the maximum possible value ($T_P' = M_P'$), while $T_*$ is the temperature of a black hole with the original Planck mass ($M_P$).

If the forms of the Uncertainty Principle and the Compton wavelength are modified to preserve duality  in the higher-dimensional case, there are two ways to generalise the above result. \\

\noindent (i) The first way assumes that $\Delta R_{3D}$ is replaced by $\Delta \mathcal{R}$ in Eq.~(\ref{3D argument}) but that $\Delta P_{3D}$ is still the relevant momentum,  this being associated with the emitted particle's rest mass. One then has 
\begin{eqnarray} 
 \Delta \mathcal{R}\simeq R_S \propto M^{1/(1+n)} , \quad \Delta P_{3D} \propto 1/( \Delta \mathcal{R})^{1+n} , \quad T_H \propto \Delta P_{3D} \propto M^{-1} \, ,
\label{dualHD}
\end{eqnarray} 
where the second relation comes from Eq.~(\ref{(3+n)-D_UR_combined***A}) and the last one is consistent with the notion that a particle can only be emitted if $T_H$ exceeds its rest mass. 
In this case, the black hole temperature reverts to the standard  Hawking expression, 
without any dependence on $n$, and the 
largest  black hole temperature  is just the maximum one allowed by the theory ($T_P$).\\

\noindent (ii) The second way identifies the temperature with the surface gravity~(\ref{HDSG}), this  not applying in the first case.  This 
corresponds to replacing Eq.~(\ref{dualHD}) with
\begin{eqnarray} 
 \Delta \mathcal{R }\simeq R_S \propto M^{1/(1+n)} , \quad \Delta P_{3D} \propto 1/( \Delta R_{3D}) \propto 1/M, \quad T_H \propto 1/ \Delta \mathcal{R} \propto M^{-1/(1+n)} \, ,
\end{eqnarray} 
where the second condition is  required for consistency with Eq.~(\ref{(3+n)-D_UR_combined***A}). This is equivalent to the surface gravity argument, since 
\begin{eqnarray} 
\kappa \propto M/R_S^{2+n} \propto \Delta \mathcal{R}^{1+n}/ \Delta \mathcal{R}^{2+n} \propto 1/\Delta  \mathcal{R} \, .
\end{eqnarray} 
so the black hole temperature is still given by Eq.~(\ref{T_H_HD}) and has a maximum value of $T_*$.\\

\noindent Since both the above arguments are heuristic, we cannot be sure which one is correct, so we allow for both possibilities below. The issue is whether one associates the black hole temperature with a length scale or a momentum scale in higher dmensions, these being inequivalent if the black hole is spherically symmetric but  the particle wave-function is not. The first argument has the attraction that black holes span the entire available temperature range; the second argument  is more consistent with  the standard higher-dimensional analysis.

We now consider the consequences of these results for PBH evaporation. In the $3$-dimensional model  ($n=0$), PBHs complete their evaporation at the present epoch if they have an initial $M_0 \simeq 10^{15}$g and an initial radius $R_0 \simeq 10^{-13}$cm, comparable to the size of a proton \cite{cksy}. For most of their lifetime these PBHs  are producing photons with energy $E_0 \simeq 100$~MeV, so the extragalactic $\gamma$-ray background at this energy places strong constraints on their number density
and current explosion rate \cite{cline}. In principle, these PBHs could also contribute to cosmic-ray positrons and antiprotons, although there are other possible sources of these particles \cite{cksy}.

However, the black holes evaporating at the present epoch are necessarily higher dimensional if $R_E > 10^{-13}$cm. In the TeV quantum gravity scenario, for example, Eq.~(\ref{nconstraint}) implies that this condition is always satisfied for $n< 7$ and this is expected  in M-theory because the maximum number of compactified dimensions is $7$. Figure~5 shows the $(M,R)$ diagram for a hierarchical  scenario with three extra dimensions, compactified on scales $R_1$, $R_2$ and $R_3$. We therefore need to recalculate the critical mass and temperature of PBHs evaporating at the present epoch, distinguishing between the standard case in which duality is broken and the alternative case in which it is preserved.

If there are $n$ extra dimensions, each with compactification scale $R_E$, and if the 
Compton wavelength has the standard form, then 
 the density of black-body radiation of temperature $T$ is
\begin{eqnarray} 
\rho_{BB} \propto 
T/ R_C(T)^{n+3}\propto T^{4+n} \, 
\label{BB1}
\end{eqnarray} 
and the black hole mass loss rate for $M<M_E$ is
\begin{eqnarray} 
dM/dt \propto 
R_S^{2+n} T_H^{4+n} \propto M^{-2/(1+n)}  \, ,
\label{massloss}
\end{eqnarray} 
where we assume that the emission is into the full ($3+n$)-dimensional space. 
This leads to a black hole lifetime
\begin{eqnarray} 
\tau \simeq  \frac{M}{dM/dt}
 \simeq \left( \frac{M}{M_P} \right)^{(3+n)/(1+n)}  \left( \frac{R_E}{R_P} \right) ^{2n/(1+n)}  t_P \, ,
\end{eqnarray} 
so the critical mass of the PBHs evaporating at the present epoch becomes
\begin{eqnarray} 
M_{crit}  \simeq 10^{15} {\rm g} \left( \frac{t_0}{t_P} \right) ^{2n/3(1+n)} \left( \frac{R_P}{R_E} \right)^{2n/(3+n)} 
\end{eqnarray} 
and the associated temperature is 
\begin{eqnarray} 
T_{crit}  \simeq 100 \, {\rm MeV} \left( \frac{t_0}{t_P} \right)^{n/3(3+n)} \left( \frac{R_P}{R_E} \right) ^{n/(3+n)} \, .
\end{eqnarray} 
Thus both $M_{crit}$ and $T_{crit}$ are modified compared to the $3$-dimensional case ($n=0$). This means that all the standard constraints on PBHs evaporating at the present epoch need to be recalculated, although we do not attempt this here. 
If there is a hierarchy of extra dimensions, the value of $n$ in the above equations must be replaced by $k$
for $R_{k+1} < R_{crit} < R_k$ ($k \leq n$). 

If T-duality is preserved, the situation is very different and there are several sources of uncertainty in modifying Eq.~(\ref{massloss}). The first concerns whether the back hole temperature is given by Eq.~(\ref{T_H}) or (\ref{T_H_HD}). The second concerns the power of $T$ in Eq.~(\ref{massloss}), which depends on the density of black-body radiation with  temperature $T$ in higher dimensions. This also relates to whether the temperature is associated with a momentum or a length scale (i.e. the first issue). Perhaps the most natural assumption is that
it is given by 
\begin{eqnarray} 
\rho_{BB} \propto 
T/ R_C(T)^{n+3}\propto T^{2(2+n)/(1+n)} \, 
\label{BB2}
\end{eqnarray} 
rather then Eq.~(\ref{BB1}),  where we have assumed $R_C \propto T^{-1/(1+n)}$ in accordance with the expression for the modifed Compton wavelength. 
However, if the   particle wave-function is non-spherical, with $R_i =R_E$ in the extra dimensions, one might expect
\begin{eqnarray} 
\rho_{BB} \propto 
T/ [R_C(T)^3 R_E^n] \propto T^4\, .
\label{BB3}
\end{eqnarray} 
The third uncertainty concerns the power of $R_S$ in Eq.~(\ref{massloss}). This is $n+2$ in the totally spherically syymmetric case. However, if  black-body particles have a scale $R_i =R_E$ in the extra dimensions, one might expect them to be confined to that scale, in which case the effective black hole area scales as $R_S^2$. 

With so many uncertainities, we cannot advocate any expression for $dM/dt$ with confidence. Only if one adopts
the combination of Eqs.~(\ref{T_H}), (\ref{BB2}) and the last argument 
does one obtain the same scaling as  in the standard 3-dimensional scenario, with the mass of PBHs evaporating today and the associated temperature preserving their standard Hawking values. 

\begin{figure}  \label{Fig5}
\begin{center}
     \psfig{file=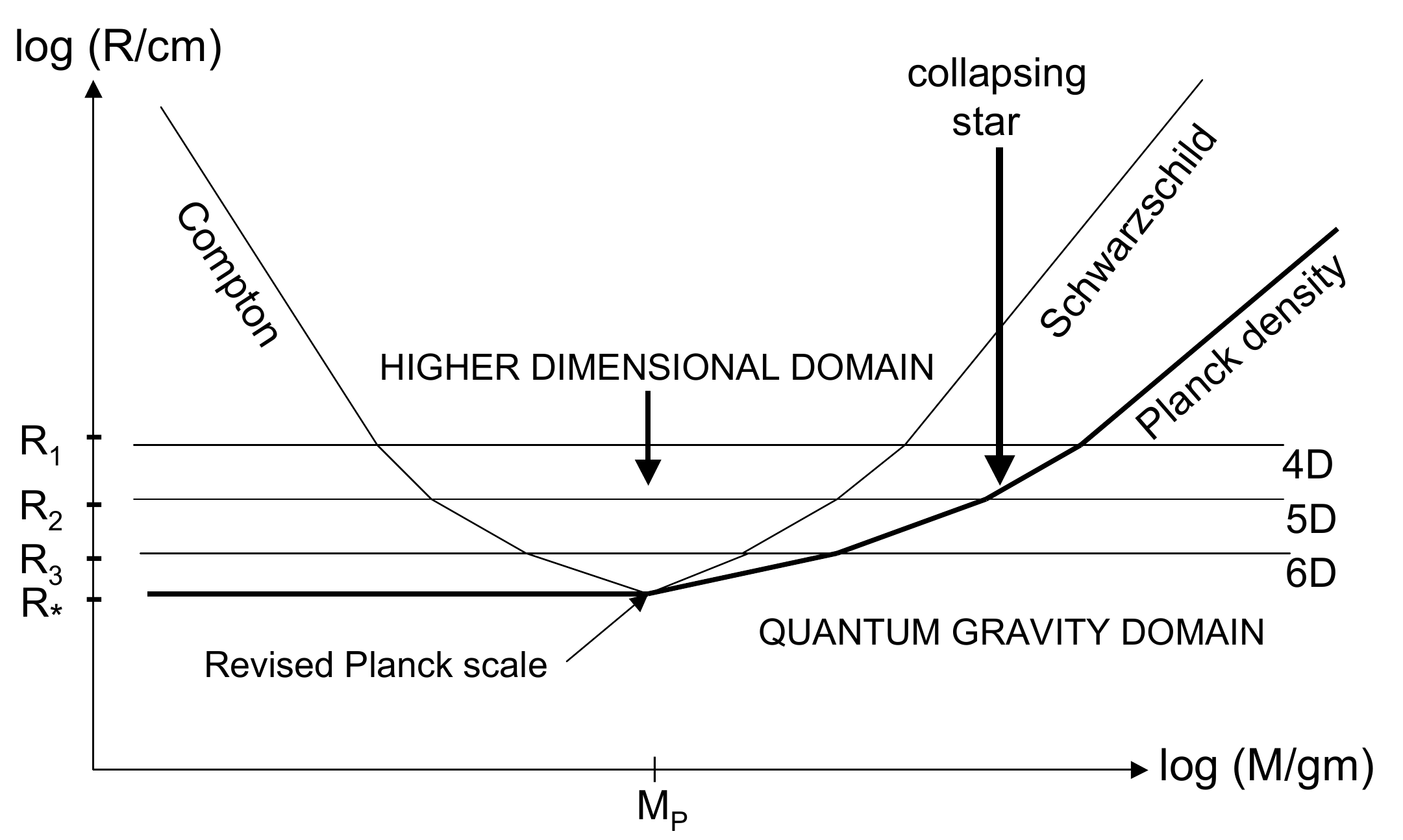,width=3.5in}
     \caption{Showing the form of the Compton and Schwarzschild scales  
for a hierarchical model with three compactified dimensions in which the Compton-Schwarzschild 
duality is preserved. In this case, the Planck length but not the Planck mass is modified and the collapsing matter may enter the quantum gravity regime at the modified Planck density. }
\end{center}
\end{figure}

\subsection{Consistency with $D$-particle scattering results}  \label{Sec.6.2}

We now consider the consistency of our phenomenologically general results with respect to the leading higher-dimensional theory of fundamental physics: string theory. In particular, we focus on their consistency with minimum-radius results for higher-dimensional, non-relativistic and quantum mechanical particle-like objects, known as $D$-particles. 
The end points of open strings obey Neumann or Dirichlet boundary conditions (or a combination of both) and are restricted to $(p+1)$-dimensional submanifolds, where $p \leq 3+n$, called $D_p$-branes. Although these are composite rather than  fundamental objects, they have dynamics in their own right and an intrinsic tension $\mathcal{T}_p = (g_sl_s^{p+1})^{-1}$, where $g_s$ denotes the string coupling and $l_s$ is the fundamental string length scale \cite{Hossenfelder:2012jw}. Thus, $D_0$-branes, also referred to as $D$-particles, are point-like, and possess internal structure only on scales $\lesssim g_sl_s$. This may be seen as the analogue of the Compton wavelength in $D_0$-brane models of fundamental particles.

At high energies, strings can convert kinetic into potential energy, thereby increasing their extension and counteracting attempts to probe smaller distances. Therefore, the best way to probe $D_p$-branes is by scattering them off each other, instead of using fundamental strings as probes \cite{Bachas:1998rg}. $D$-particle scattering has been studied in detail by Douglas {\it et al} \cite{Douglas:1996yp}, who showed that slow moving $D$-particles can be used to probe distances down to $g_s^{1/3}l_s$ in $D=10$ spacetime dimensions. 

This result may be obtained heuristically as follows \cite{Hossenfelder:2012jw}. Let us consider a perturbation of the metric component $g_{00} = 1+2V$ induced by the Newtonian potential $V$ of a higher-dimensional particle of mass $M$. In $D$ spacetime dimensions, this takes the form
\begin{eqnarray} \label{Dpart-1}
V \simeq -\frac{G_DM}{(\Delta x)^{D-3}} \, ,
\end{eqnarray}
where $\Delta x$ is the spatial extension of the particle and $G_D$ is the $D$-dimensional Newton's constant, so that the horizon is located at
\begin{eqnarray} \label{Dpart-2}
\Delta x \simeq (G_DM)^{1/(D-3)}. 
\end{eqnarray}
(For convenience, we set $c=\hbar=1$ throughout this section.) In spacetimes with $n$ compact spatial dimensions, this is related to the $(3+1)$-dimensional Newton's constant via $G \simeq G_D/R_E^n$, so that, for $D = 3+n+1$, we simply recover the formula for the higher-dimensional Schwarzschild radius (\ref{higherBH}).

However, we may also use Eq.~(\ref{Dpart-2}) to derive the minimum length obtained from $D$-particle scattering in \cite{Douglas:1996yp} by first setting $M \simeq 1/\Delta t$, where $\Delta t$ is the time taken to test the geometry, and then using the higher-dimensional Newton's constant derived from string theory, $G_D \simeq g_s^2 l_s^{D-2}$ \cite{Zwiebach:2004tj}. This gives
\begin{eqnarray} \label{Dpart-3}
(\Delta t)(\Delta x)^{D-3} \gtrsim g_s^2l_s^{D-2} \, .
\end{eqnarray}
Combining this with the spacetime uncertainty principle, which is thought to arise as a consequence of the conformal symmetry of the $D_p$-brane world-volume \cite{Yoneya:1989ai,Yoneya:2000bt,Jevicki:1998yr},
\begin{eqnarray} \label{Dpart-4}
\Delta x \Delta t \gtrsim l_s^2,
\end{eqnarray}
we then have
\begin{eqnarray} \label{Dpart-5}
(\Delta x)_{\rm min} \simeq g_s^{2/(D-4)}l_s, \ \ \ (\Delta t)_{\rm min} \simeq g_s^{-2/(D-4)}l_s.
\end{eqnarray}
For $D=10$, this gives $(\Delta x)_{\rm min} \sim g_s^{1/3}l_s$, as claimed.
 
Combining results from string theory and higher-dimensional general relativity by setting $G_D \simeq g_s^2l_s^{D-2} \simeq R_{P}^2R_E^{D-4}$ with $D=3+n+1$, we obtain
\begin{eqnarray} \label{Dpart-6}
R_{P}' \simeq (R_{P}^2R_E^n)^{1/(2+n)} \simeq g_s^{2/(2+n)}l_s \, ,
\end{eqnarray}
which gives $R_{P}' \simeq g_s^{1/4}l_s$ as the modified Planck length for $D = 10$. In fact, Eqs. (\ref{Dpart-5})-(\ref{Dpart-6}) suggest that the minimum positional uncertainty for $D$-particles {\it cannot} be identified with the modified Planck length obtained from the intersection of the higher-dimensional Schwarzschild line, $R_S \sim M^{1/(1+n)}$, and the standard Compton line, $R_C \sim M^{-1}$, in any number of dimensions. Hence, the standard scenario is {\it incompatible} with $D$-particle scattering results.

However, it is straightforward to verify that, if $R_{P} \simeq g_s^{-2/n}l_s$, we have $R_* \simeq (R_{P}R_E^n)^{1/(1+n)} \simeq (\Delta x)_{\rm min} \simeq g_s^{2/n}l_s$, so that the intersection of the higher-dimensional Schwarzschild and Compton lines is equal to the minimum length scale that can be probed by $D$-particles. In this scenario, $R_E \simeq g_s^{2(1+n)/n^2}l_s$, and we note that $R_E \rightarrow R_* \rightarrow R_{P}' \rightarrow R_{P} \rightarrow l_s$ for $g_s \rightarrow 1$, as required for consistency. In general, $R_* > R_{P}'$ (or equivalently $R_E > R_{P}$) requires $g_s > 1$.

\section{Conclusions} \label{Sec.7} 

We have addressed the question of how the {\it effective} Compton wavelength of a fundamental particle -- defined as the minimum possible positional uncertainty over measurements in {\it all} independent spatial directions  -- scales with mass if there exist $n$ extra compact dimensions. In $(3+1)$-dimensional spacetime, the Compton wavelength scales as $R_C \sim M^{-1}$, whereas the Schwarzschild radius scales as $R_S \sim M$, so the two are related via $R_S \sim R_{P}^2/R_C$. In higher-dimensional spacetimes with $n$ compact extra dimensions, $R_S \sim M^{1/(1+n)}$ on scales smaller than the compactification radius $R_E$, which breaks the symmetry between particles and black holes if the Compton scale remains unchanged. However, we have argued that the effective Compton scale depends on the form of the wavefunction in the higher-dimensional space. If this is spherically symmetric in the three large dimensions, but maximally asymmetric in the full $3+n$ spatial dimensions, then the effective radius scales as $R_C \sim M^{-1/(1+n)}$ rather than $M^{-1}$ on scales less than $R_E$ and this preserves the symmetry about the $M \simeq M_{P}$ line in ($M,R$) space. 

In this scenario, the effective Planck length is increased but the Planck mass is unchanged, so quantum gravity and microscopic black hole production are associated with the standard Planck energy, as in the 3-dimensional scenario. On the other hand, one has the interesting prediction that the Compton line -- which marks the onset of pair-production -- is ``lifted", relative to the 3-dimensional case, in the range $R_{P} < R < R_E$, so that extra-dimensional effects may become visible via enhanced pair-production rates for particles with energies $E > M_Ec^2 = \hbar c/R_E$. This prediction is consistent with minimum length uncertainty relations obtained from $D$-particle scattering amplitudes in string theory.  
Also, as indicated in Fig.~4, the existence of extra compact dimensions has crucial implications for the detectability of black holes evaporating at the present epoch, since they are {\it necessarily} higher-dimensional for $R_E > 10^{-13}$cm.

In this paper, we have assumed that non-relativistic quantum mechanical particles obey the standard Heisenberg Uncertainty Principle (HUP) in each spatial direction. The modified expression for the {\it effective} Compton line, which retains a simple power-law form until its intersection with the higher-dimensional Schwarzschild line in the ($M,R$) diagram, is seen to arise from the application of the HUP to maximally asymmetric wave functions. These are spherically symmetric with respect to the three large dimensions but pancaked in the compact directions. No allowance has been made for deviations from the HUP, as postulated by various forms of Generalised Uncertainty Principle (GUP) proposed in the quantum gravity literature, and no attempt has been made to smooth out the transition between particle and black hole states at the Planck point, as postulated by the Black Hole Uncertainty Principle (BHUP) correspondence \cite{Ca:2014}. 
These effects would entail different temperature predictions in the Planck regime even in the $3$-dimensional case. Our main intention here has been to examine the consequences of the existence of extra dimensions in the `standard' (i.e. HUP-based) scenario.
Many other authors have studied the implication of the GUP for higher dimensional models \cite{koppel} but without imposing (semi-)T-duality.

Finally, if we interpret the Compton wavelength as marking the boundary on the $(M,R$) diagram below which pair-production rates becomes significant, we expect the presence of compact extra dimensions to affect pair-production rates at high energies. Specifically, we expect pair-production rates at energies above the 
mass scale associated with the compact space, $M_E \equiv \hbar/(cR_E)$, to be enhanced relative to the 3-dimensional case. This is equivalent to raising the Compton line, i.e. increasing its (negative) gradient in the $(M,R)$ diagram. A more detailed relativistic analysis would be needed to confirm whether this is a generic result for massive scalar fields (corresponding to uncharged matter). There is tentative theoretical evidence that enhanced pair-production may be a generic feature of higher-dimensional theories in which some directions are compactified
but the available literature on this is sparse (c.f. \cite{He99,Ebetal00}).

\begin{center}
{\bf Acknowledgments}
\end{center}

BC thanks the Research Center for the Early Universe (RESCEU), University of Tokyo, and ML thanks the Institute for Fundamental Study at Naresuan University 
for hospitality received during this work. We thank John Barrow, Tiberiu Harko, Juan Maldacena, Shingo Takeuchi, Pichet Vanichchapongjaroen and Marek Miller for helpful comments and discussions.

\appendix

\section{Intepretation of the uncertainty principle} \label{Sec.A}

In the form originally derived by Heisenberg, the uncertainty principle states that the product of the ``uncertainties" in the position and momentum of a quantum mechanical particle is of order of or greater than the reduced Planck's constant $\hbar$ \cite{He27}. More generally, the rigorous definition of the uncertainty $\Delta_\psi O$ for an operator $\hat{O}$ is the standard deviation for a large number $N$ of (absolutely precise) repeated measurements  of an ensemble of identically prepared systems described by the wave vector $\Ket{\psi}$: 
\begin{eqnarray} \label{sd}
\Delta_\psi  O = \sqrt{\langle \psi|\hat{O}^2|\psi\rangle - \langle \psi|\hat{O}|\psi\rangle^2} \, .
\end{eqnarray}
Formally, this expression corresponds to the limit $N \rightarrow \infty$ and is generally $|\psi\rangle$-dependent. Thus the uncertainty $\Delta_\psi O$ does not correspond to incomplete knowledge about the value of the property $O$ for the system, since $\Ket{\psi}$ need not possess a definite value of $O$. 

Consistency with the Hilbert space structure of quantum mechanics requires that the product of the uncertainties associated with arbitrary operators $\hat{O}_1$ and $\hat{O}_2$ satisfy the bound  \cite{Rae00,Ish95}
\begin{eqnarray} \label{SUP}
\Delta_\psi  O_1 \Delta_\psi  O_2 &\geq& \frac{1}{2}\sqrt{|\langle \psi|[\hat{O}_1,\hat{O}_1]|\psi\rangle|^2 + |\langle \psi|[\hat{A},\hat{B}]_{+}|\psi\rangle|^2}
\geq \frac{1}{2}|\langle \psi|[\hat{O}_1,\hat{O}_1]|\psi\rangle| \, ,
\end{eqnarray}
where $[\hat{O}_1,\hat{O}_2]$ is the commutator of $\hat{O}_1$ and $\hat{O}_2$ and $[\hat{A},\hat{B}]_{+}$ is the anticommutator of $\hat{A} = \hat{O}_1 - \langle \hat{O}_1\rangle\hat{\mathbb{I}}$ and $\hat{B} = \hat{O}_2 - \langle \hat{O}_2\rangle\hat{\mathbb{I}}$. This formulation, which was first presented in Refs.~\cite{Ro29,Sc30}, can also be given a measurement-independent interpretation since, from a purely mathematical perspective, $\Delta_\psi  O_1$ and $\Delta_\psi  O_2$ represent the ``widths" of the wave function in the relevant physical space or phase space, regardless of whether a measurement is actually performed.

For the operators $\hat{x}$ and $\hat{p}_x$, defined by $\hat{x}\psi(x)=x\psi(x)$ and $\hat{p}_x\psi(p_x)=p_x\psi(p_x)$, the commutation relation 
$[\hat{x},\hat{p}_x] = i\hbar$ gives
\begin{eqnarray} \label{SUP_xp}
\Delta_\psi  x\Delta_\psi  p_x \geq \hbar/2 \, ,
\end{eqnarray}
where $\Delta_\psi  x$ and $\Delta_\psi  p_x$ correspond to the standard deviations of $\psi(x)$ in position space and $\psi(p_x)$ in momentum space, respectively. This formulation of the uncertainty principle for $\hat{x}$ and $\hat{p}_x$ was first given in Refs.~\cite{Ke27,We28} and, for this choice of operators, the $|\psi\rangle$-dependent terms in Eq. (\ref{SUP}) are of subleading order, in accordance with Heisenberg's original result. The underlying wave-vector in the Hilbert space of the theory is identical in either the physical or momentum space representations, which correspond to different choices for the basis vectors in the expansion of $|\psi\rangle$ \cite{Rae00,Ish95}. 

Although $\Delta_\psi  x$ and $\Delta_\psi  p_x$ do \emph{not} refer to any unavoidable ``noise", ``error" or  ``disturbance" introduced into the system by the measurement process, this was how Heisenberg interpreted his original result \cite{He27}. In order to distinguish between quantities representing such noise and the standard deviation of repeated measurements which do not disturb the state $\ket{\psi}$ prior to wave function collapse, within this Appendix (but not the main text) we use the notation $\Delta O$ for the former and $\Delta_\psi O$ for the latter. 
Strictly speaking, any disturbance to the state of the system caused by an act of measurement may also be $|\psi\rangle$-dependent, but we adopt 
Heisenberg's original notation, in which the state-dependent nature of the disturbance is not explicit.
In this notation, Heisenberg's original formulation of the uncertainty principle may be written as
\begin{eqnarray} \label{HUP_xp}
\Delta x\Delta p_x \gtrsim \hbar \ ,
\end{eqnarray}
ignoring numerical factors. It is well known that one can heuristically understand this result as reflecting the momentum transferred to the particle by a probing photon. However, such a statement must be viewed as a postulate, with no rigorous foundation in the underlying mathematical structure of quantum theory. Indeed, as a postulate, it has been shown to be manifestly false, both theoretically \cite{Oz03A,Oz03B} and experimentally \cite{Roetal12,Eretal12,Suetal13,Baetal13}. 

Despite this, the heuristic derivation of Eq. (\ref{HUP_xp}) may be found in many older texts, alongside the more rigorous derivation of Eq. (\ref{SUP}) from basic mathematical principles (see, for example, \cite{Rae00}). Unfortunately, it is not always made clear that the quantities involved in each expression are different, as clarified by the pioneering work of Ozawa \cite{Oz03A,Oz03B}. An excellent discussion of the various possible meanings and (often confused) interpretations of symbols like `$\Delta x$' is given in \cite{Scheibe}. 
We consider only uncertainties of the form $\Delta_\psi O$, defined in Eq.~(\ref{sd}), and uncertainty relations derived from the general formula Eq.~(\ref{SUP}). However, for notational convenience we do not include the subscript $\psi$ in the main text.
Unfortunately, Eq.~(\ref{SUP}) is also sometimes referred to as the Generalized Uncertainty Principle or Generalized Uncertainty Relation (see, for example, \cite{Ish95}). To avoid confusion, we use the term {\it General} Uncertainty Principle to refer to the most general uncertainty relation obtained from the Hilbert space structure of standard non-relativistic quantum mechanics (for arbitrary operators) and the term {\it Generalized} Uncertainty Principle to refer to the amended uncertainty relation for position and momentum in non-canonical theories. 

\section{Pair-production in the non-spherical case} \label{Sec.B}

For collisions between pairs of non-relativistic free particles in momentum eigenstates with masses $M$ and $M'$, pair-production of particles with rest mass $M$ is possible if the centre-of-mass frame energy satisfies 
\begin{eqnarray} \label{pair-prod-1}
E \simeq \frac{P_{\rm 3D}^2}{2\mu} 
\gtrsim Mc^2 \, , 
\end{eqnarray}
where $\mu = MM'/(M+M')$ is the reduced mass.
(Reversing the direction of the final equality is the condition for {\it non}-pair-production; all the inequalities below can be similarly negated but we will not state this explicitly.) Here $P_{\rm 3D}$ denotes the 3-momentum of each particle
and their total 3-momentum 
in the centre-of-mass frame is zero by definition. For identical particles, $M' = M$, so $\mu =M/2$ and Eq. (\ref{pair-prod-1}) reduces to
\begin{eqnarray} \label{pair-prod-2}
P_{\rm 3D}^2 
= \hbar^2k^2 = \hbar^2(k_x^2+k_y^2+k_z^2) 
 = h^2\left(\frac{1}{\lambda_x^2}+\frac{1}{\lambda_y^2}+\frac{1}{\lambda_z^2} \right) 
\gtrsim  M^2c^2 \, 
\end{eqnarray}
in three (infinite) spatial dimensions. 
This may be written as
\begin{eqnarray} \label{lambda}
\lambda_{\rm 3D} \equiv \frac{\lambda_x\lambda_y\lambda_z}{\sqrt{\lambda_x^2\lambda_y^2+\lambda_x^2\lambda_z^2+\lambda_y^2\lambda_z^2}}  \lesssim R_{\rm C} = 
\frac{h}{Mc} \, ,
\end{eqnarray}
where in this Appendix $R_C$ is always the standard Compton expression. To within numerical factors of order unity, the final expression on the right-hand side of this equation remains valid when the particles have very different rest masses ($M \ll M' \Rightarrow \mu \approx M$) and when higher order relativistic effects are included in Eq.~(\ref{pair-prod-1}). 

For spherically symmetric states, $\lambda_x = \lambda_y = \lambda_z = \lambda_R $, giving 
$\lambda_{\rm 3D} = \lambda_{\rm R}/\sqrt{3} 
\lesssim R_{\rm C}$ and the volume required for pair-production is just $V_{\rm min} \simeq R_{\rm C}^3$. However, for highly asymmetric states, 
the minimum volume 
required for pair-production 
may be much larger than this, so
 the effective ``width" of the wave-packet, averaged over all dimensions, may 
far exceed $R_{\rm C}$. More specifically, if $\lambda_y \simeq \lambda_z \equiv \lambda_{\rm 2D}$, we may have spindles with 
$\lambda_x \gg \lambda_{\rm 2D}$ or pancakes with $\lambda_x \ll  \lambda_{\rm 2D}$. 
In these cases, we have
\begin{eqnarray}
\lambda_{\rm 3D} = {\rm min} (\lambda_{2D},\lambda_x) =
\begin{cases}
\lambda_{\rm 2D} 
&(\lambda_x \gg \lambda_{\rm 2D}) \\
\lambda_x 
&(\lambda_x \ll \lambda_{\rm 2D})
\end{cases}
\end{eqnarray}
and this must less than $R_C$ for pair-production. The threshold volume for this is
\begin{eqnarray}
V_{\rm min} = \lambda_x \lambda_{\rm 2D}^2 \simeq
\begin{cases}
R_C^2 \lambda_x 
&(\lambda_x \gg \lambda_{\rm 2D}) \\
R_C \lambda_{\rm 2D}^2
&(\lambda_x \ll \lambda_{\rm 2D})\, ,
\end{cases}
\end{eqnarray}
with both expressions exceeding $R_C^3$.
This may be contrasted with {\it classical} systems on the right-hand side of Fig. 1, for which $V \lesssim 
R_{\rm S}^3$ is required for gravitational collapse. 

Similar considerations apply in the presence of extra dimensions.
In $3+n$ dimensions, the total ($3+n$)-momentum may be decomposed into the 3-dimensional and extra-dimensional parts. If the extra dimensions are large (or infinite), the condition for 
pair-production becomes 
\begin{eqnarray} \label{pair-prod-HD-1}
P_T^2 = P_{\rm 3D}^2 + P_{E}^2 
\gtrsim M^2c^2 \, ,
\end{eqnarray}
where 
\begin{eqnarray} \label{pE}
P_{\rm 3D}^2 = p_x^2+p_y^2+p_z^2 \, , \quad P_{E}^2 \equiv \sum_{i=1}^{n} p_{i}^2 \, .
\end{eqnarray}
In this case, it is clear that the threshold for pair-production may be reached by increasing the momentum of the particle in either the 3-dimensional 
or extra-dimensional space or both.

The pair-production condition is changed in a non-trivial way if some
of the extra dimensions are compact.
 In the compact directions, 
the $i^{\rm th}$ component of the de Broglie wavelength
 is bounded from above by the corresponding compactification scale $R_i$, so
\begin{eqnarray} \label{lambda_i>R_i}
\lambda_{i} \lesssim R_{i} \, .
\end{eqnarray}
This gives 
 a lower bound on the $i^{\rm th}$ extra-dimensional momentum component, 
\begin{eqnarray} \label{p_i>h/M_i}
p_{i} \gtrsim M_{i}c \equiv \frac{\hbar}{R_{i}} \, ,
\end{eqnarray}
which corresponds to the minimum-energy, space-filling ground state of the particle. Since any newly created particle must also posses the minimum momentum in the compact space, the condition for 
pair-production becomes
\begin{eqnarray} \label{pair-prod-HD-2}
P_T^2 = P_{\rm 3D}^2 + P_{E}^2  
\gtrsim M^2c^2 + \mathcal{M}_{E}^2c^2 \, ,
\end{eqnarray}
where
\begin{eqnarray} \label{ME}
\mathcal{M}_{E}^2 \equiv \sum_{i=1}^{n} M_{i}^2 \equiv \frac{\hbar^2}{\mathcal{R}_E^2}  \, .
\end{eqnarray}
This condition can be written as 
\begin{eqnarray} \label{pp}
\lambda \lesssim  \frac{h}{c \sqrt{M^2 + \mathcal{M}_E^2}} \simeq
\begin{cases}
R_C
&(M \gtrsim \mathcal{M}_E) \\
\mathcal{R}_E
&(M \lesssim \mathcal{M}_E) \, ,
\end{cases}
\end{eqnarray}
where
\begin{eqnarray} \label{lambda_defns-4}
\lambda \equiv \lambda_{x}\lambda_{y}\lambda_{z} \prod_{i=1}^{n}\lambda_{i} \times \Bigg[\left(\lambda_{x}^2\lambda_{y}^2 + \lambda_{x}^2\lambda_{z}^2 + \lambda_{y}^2\lambda_{z}^2\right)\prod_{i=1}^{n}\lambda_{i}^2 + \lambda_{x}^2 \lambda_{y}^2 \lambda_{z}^2 \sum_{j=1}^{n}\prod_{i \neq j}\lambda_{i}^2 \Bigg]^{-1/2} \, 
\end{eqnarray}
is the higher-dimensional generalisation of the quantity $\lambda_{\rm 3D}$ defined by Eq.~(\ref{lambda}).
For quasi-symmetric states, corresponding to $ \lambda_{x} = \lambda_{y} = \lambda_{z} = \lambda_{R} = \sqrt{3} \, \lambda_{\rm 3D}$ 
this becomes 
\begin{eqnarray} \label{lambda_defns-4*}
\lambda \equiv \lambda_{\rm 3D}\prod_{i=1}^{n}\lambda_{i} \times \Bigg[\prod_{i=1}^{n}\lambda_{i}^2 + \lambda_{\rm 3D}^2 \sum_{j=1}^{n}\prod_{i \neq j}\lambda_{i}^2 \Bigg]^{-1/2} \, .
\end{eqnarray}
For spindle  configurations with $\lambda_i \gg \lambda_{\rm 3D}$ and pancake configurations with  $\lambda_i \ll \lambda_{\rm 3D}$, we have
\begin{eqnarray}
\lambda =
\begin{cases}
\lambda_{\rm 3D} 
&(\lambda_i \gg \lambda_{\rm 3D}) \\
\prod_{i=1}^{n}\lambda_{i}/(\sum_{j=1}^{n}\prod_{i \neq j}\lambda_{i}^2)^{1/2}
&(\lambda_i \ll \lambda_{\rm 3D}) \, .
\end{cases}
\end{eqnarray} 
The threshold volume for pair-production is
\begin{eqnarray}
V_{\rm min} = \lambda_{\rm 3D}^3 \prod_{i=1}^{n}\lambda_{i} \simeq
\begin{cases}
R_C^3 \prod_{i=1}^{n} \lambda_{i}
&(\lambda_i \gg \lambda_{\rm 3D}) \\
\lambda_{\rm 3D}^3 R_C^{n} 
&(\lambda_i \ll \lambda_{\rm 3D})\, ,
\end{cases}
\end{eqnarray}
both potentially exceeding $R_C^{n+3}$. However, this is not the {\it key} quantity controlling pair-production.

Rewriting the right-hand side of Eq.~(\ref{pp}) in terms of $R_C$ and $R_i$, the pair-production condition for quasi-symmetric states can be written as
\begin{eqnarray} \label{Delta_lambda-1}
 \lambda^{1+n} 
\equiv \lambda_{\rm 3D}\prod_{i=1}^{n}\lambda_i 
\lesssim R_{\rm C} \prod_{i=1}^{n}R_{i} 
\times \Bigg[ \frac{\prod_{i=1}^{n}\lambda_i^2 + \lambda_{\rm 3D}^2 \sum_{j=1}^{n}\prod_{i \neq j}\lambda_i^2}{ 
\prod_{i=1}^{n}R_{i}^2 + R_{C}^2\sum_{j=1}^{n}\prod_{i \neq j}R_{i}^2 }\Bigg]^{1/2} \, . 
\end{eqnarray}
For the pancake configurations corresponding to the experimental scenarios outlined in Sec.~5, one expects $\lambda_{\rm 3D} \lesssim R_{C} \lesssim \lambda_i \lesssim R_i$ for all $i$,
so  the term in square brackets is less than $1$, which implies
\begin{eqnarray} \label{Delta_lambda-2}
 \lambda^{1+n} \equiv \lambda_{\rm 3D}\prod_{i=1}^{n}\lambda_i  
\lesssim R_{C} \prod_{i=1}^{n}R_{i} \, . 
\end{eqnarray}
Since the last expression yields the threshold value of $\lambda$ giving rise to pair-production, it represents a minimum width for the particle. The critical limiting value on the right-hand side of Eq.~(\ref{Delta_lambda-2}) is reached from below with respect to $\lambda_i$ (i.e. as $\lambda_i \rightarrow R_i^{-}$) but from above with respect to $\lambda_{\rm 3D}$ (i.e. as $\lambda_{\rm 3D}\rightarrow R_{C}^{+}$). States with $\lambda_i \simeq R_i$, where the volume of the wave function in the extra dimensions remains 
as large as possible, represent the {\it maximal} degree of asymmetry. More generally, for particles that are not in momentum eigenstates, we may 
put $\lambda_{\rm 3D} \rightarrow \Delta R_{\rm 3D}$, $\lambda_{i} \rightarrow \Delta x_{i}$ and  $\lambda^n \rightarrow (\Delta \mathcal{R})^{1+n}$ in Eq.~(\ref{Delta_lambda-2}). Analogous arguments to those given above then lead to
\begin{eqnarray} \label{Delta_lambda-2*}
(\Delta \mathcal{R})^{1+n} \equiv \Delta R_{\rm 3D}\prod_{i=1}^{n}\Delta x_{i} 
\lesssim R_{C} \prod_{i=1}^{n}R_{i} \, ,
\end{eqnarray}
which is the converse of the (non-pair-production) condition (\ref{(3+n)-D_UR*1}). 
The right-hand side of Eq.~(\ref{Delta_lambda-2*}) equals $R_*$, given by  
Eq.~(\ref{R_*}), when $R_i = R_{E}$ for all $i$.

\end{document}